\documentclass[lettersize,journal]{IEEEtran}
\usepackage{amsmath,amsfonts}
\usepackage{algorithmic}
\usepackage{algorithm, multirow}
\usepackage{array}
\usepackage[caption=false,font=normalsize,labelfont=sf,textfont=sf]{subfig}
\usepackage{textcomp}
\usepackage{stfloats}
\usepackage{url}
\usepackage{verbatim}
\usepackage{graphicx, marginnote}
\usepackage{cite,color}
\begin{document}

\title{Set-Theoretic Receding Horizon Control for Obstacle Avoidance and Overtaking in Autonomous Highway Driving}

\author{Gianni Cario, Valentino Carriuolo, Alessandro Casavola, Gianfranco Gagliardi*, Marco Lupia, Franco Angelo Torchiaro
        % <-this % stops a space
\thanks{The authors are within the Dipartimento di Ingegneria Informatica, Modellistica, Elettronica e Sistemistica (DIMES), Universit\`{a} della Calabria, 87046, Rende (CS), Italy}% <-this % stops a space
\thanks{Manuscript received  XXXXX, XXXX; revised XXXXXX XX, XXXX.}}

\markboth{Journal of \LaTeX\ Class Files,~Vol.~XX, No.~X, XXXXX~XXXX}%
{Cario \MakeLowercase{\textit{et al.}}: Set-Theoretic Receding Horizon Control for Obstacle Avoidance and Overtaking in Autonomous Highway Driving}


\maketitle
\begin{abstract}
This article addresses obstacle avoidance motion planning for autonomous vehicles, specifically focusing on highway overtaking maneuvers. The control design challenge is handled by considering a mathematical vehicle model that captures both lateral and longitudinal dynamics. 
Unlike existing numerical optimization methods that suffer from significant online computational overhead, this work extends the state-of-the-art by leveraging a fast set-theoretic ellipsoidal Model Predictive Control (Fast-MPC) technique. While originally restricted to stabilization tasks, the proposed framework is successfully adapted to handle motion planning for vehicles modeled as uncertain polytopic discrete-time linear systems.
%\textcolor[rgb]{1,0,0}{Unlike existing numerical optimization approaches that suffer from high online computational burdens, this work advances the state-of-the-art by extending a fast set-theoretical ellipsoidal Model Predictive Control (Fast-MPC) technique. While previously limited to stabilization tasks, the approach is successfully extended to handle motion planning for vehicles modeled as uncertain polytopic discrete-time linear systems.}
The control action is computed online via a set-membership evaluation against a structured sequence of nested inner ellipsoidal approximations of the exact one-step ahead controllable set within a receding horizon framework. A six-degrees-of-freedom (6-DOF) nonlinear model characterizes the vehicle dynamics, while a polytopic embedding approximates the nonlinearities within a linear framework with parameter uncertainties. 
Finally, to assess performance and real-time feasibility, comparative co-simulations against a baseline Non-Linear MPC (NLMPC) were conducted. Using the high-fidelity CARLA 3D simulator, results demonstrate that the proposed approach seamlessly rejects dynamic traffic disturbances while reducing online computational time by over 90\% compared to standard optimization-based approaches.
\end{abstract}


\section{Introduction}
\label{s0}
\begin{figure}[htbp]
	\centering
		\includegraphics[width=0.5\textwidth]{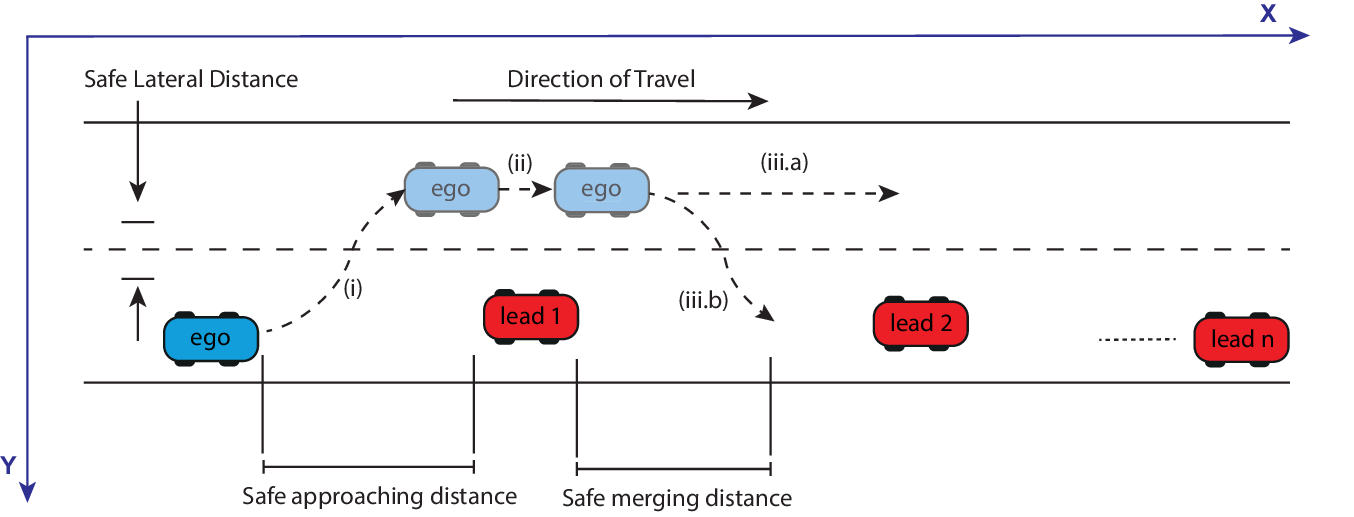}
	\caption{Overtaking maneuver: the autonomous (ego) vehicle executes a lane change (i) to overtake the preceding (lead) vehicle (ii). After successfully passing the lead vehicle, the ego vehicle either returns to its original lane (iii.a) or continues in the new lane if necessary (iii.b), depending on traffic conditions. (\textbf{X},\textbf{Y}) global reference frame}
	\label{fig:overtaking}
\end{figure}

Driven by increasingly stringent traffic safety requirements, Advanced Driver Assistance Systems (ADAS) \cite{r1} and Intelligent Autonomous Vehicles (IAVs) \cite{r2} rely heavily on robust environmental perception and motion planning. To autonomously regulate steering, braking, and speed, these vehicles must accurately process real-time traffic data acquired through onboard sensors (e.g., cameras, radar, LiDAR) \cite{r3} and V2X communication.
%In the last decade, cars have evolved into advanced technological systems designed to enhance passenger comfort, improve fuel efficiency, reduce emissions, and, more broadly, meet increasingly stringent traffic safety requirements.
%To this end, research on Advanced Driver Assistance Systems (ADAS) \cite{r1} and Intelligent Autonomous Vehicles (IAV) has gained significant attention, driven by technological advancements in sensing, communication, and information processing \cite{r2}. An intelligent or autonomous vehicle is defined as a vehicle that performs certain driving tasks either autonomously or by assisting the driver in performing them more effectively. The design of an autonomous vehicle involves various challenges, including environmental perception, motion planning, vehicle positioning, and path tracking. Consequently, the vehicle must accurately acquire motion parameters and real-time traffic information using onboard sensors (e.g., cameras, radar, LiDAR) \cite{r3} and V2X communication, which are processed to regulate steering, braking, and speed. 
Obstacle Avoidance Motion Planning (OAMP) remains a fundamental challenge for autonomous vehicles in dynamic environments with a priori unknown obstacles \cite{[R4],r32}. Such scenarios demand adaptive, real-time control strategies to ensure safe trajectory tracking and strict constraint satisfaction \cite{r28,r32}.
%Obstacle Avoidance Motion Planning (OAMP) and control represent fundamental challenges in the field of autonomous vehicles. This problem entails the determination of an optimal sequence of control inputs that enables the vehicle to track a predefined trajectory while satisfying a set of predefined constraints. In many real-world applications, autonomous vehicles must operate in dynamic environments with a priori unknown obstacles, requiring adaptive and real-time motion planning strategies.
While obstacle avoidance has been explored in diverse settings—ranging from 3D trajectory tracking for multi-quadrotor systems \cite{Ref_Quadrotor} to navigating 'negative obstacles' (e.g., potholes) for ground vehicles \cite{Ref_NegativeObs}—high-speed highway driving presents unique challenges. These scenarios underscore the need for robust, real-time control architectures capable of managing severe uncertainties. Within this framework, our work focuses on high-speed overtaking maneuvers, where maintaining a balance between agility and safety is paramount.
Overtaking is a critical obstacle avoidance maneuver in autonomous driving. Because its execution depends heavily on dynamic traffic conditions and surrounding obstacles, ensuring a safe and efficient maneuver requires the simultaneous control of both longitudinal and lateral vehicle dynamics, guided by real-time sensor data.
%Overtaking is one of the most common obstacle avoidance maneuvers in autonomous driving. In this context, an autonomous vehicle must determine whether, when, and how to execute this maneuver while ensuring safety, efficiency, and compliance with traffic regulations. The overtaking maneuver is a complex task, as it is not a standardized procedure and depends on various external factors, including traffic conditions, vehicle speed, surrounding obstacles, and road regulations. To ensure a safe and efficient maneuver, both the longitudinal and lateral dynamics of the vehicle must be considered, along with real-time sensor data for an accurate interpretation of the surrounding environment.
As illustrated in Figure \ref{fig:overtaking}, the overtaking maneuver typically involves three phases: lane change, passing, and returning to the original lane.
Executing this maneuver requires robust trajectory planning and tracking, especially at high speeds where stability depends on precise dynamic control across the lane-changing and passing phases.

%Two critical functions in this maneuver are trajectory planning and trajectory tracking, both of which are essential for executing a smooth and collision-free overtaking action. Additionally, speed plays a crucial role, as high-speed maneuvers require precise knowledge of both vehicle dynamics and environmental conditions to maintain stability and safety.
%As illustrated in Figure \ref{fig:overtaking}, the overtaking maneuver typically involves three phases: lane change, passing, and returning to the original lane.
%The overtaking maneuver can be generally divided into three phases, as illustrated in the schematic representation in the Figure \ref{fig:overtaking}:(i) lane change; (ii) pass front vehicle; (iii) lane change back to original lane or continue the overtaken maneuver.
%
Recent literature addresses autonomous overtaking primarily through AI-based methods and mathematical control techniques. Data-driven approaches - including Machine Learning, Deep Reinforcement Learning (DRL), and fuzzy systems - have proven highly effective in directly mapping complex inputs to overtaking actions \cite{r8}-\cite{r18}.
%A number of approaches have been proposed to improve both safety and efficiency in trajectory planning and tracking, and various approaches (e.g. AI-based solution, mathematical control techniques, etc.) are used to solve the overtaking issues considering different scenarios to validate the methods in simulations and real-life scenarios. AI-based techniques rely on end-to-end processing approaches that directly map inputs to outputs. Methods such as deep learning (DL), machine learning (ML), reinforcement learning (RL), deep reinforcement learning (DRL), and fuzzy systems have proven effective in executing overtaking maneuvers \cite{r8}-\cite{r18}. 

%%% DA QUI
On the other side, mathematical control approaches involve modeling various scenarios and translating vehicle dynamics into mathematical formulations, often used to enable overtaking maneuvers. Notable methods include standard, stochastic, and mixed-integer MPC \cite{r19}-\cite{r22}, \cite{r25}, alongside sliding mode control \cite{r24} (see \cite{r27} for a comprehensive survey). 
Despite their effectiveness, such approaches frequently encounter limitations in balancing real-time performance and robustness against dynamic uncertainties, especially when strict constraint satisfaction is required.
%While these methods offer valuable solutions, they frequently face challenges in simultaneously ensuring real-time computational efficiency and robust constraint satisfaction under dynamic uncertainties.}
%
%On the other side, mathematical control approaches involve modeling various scenarios and translating vehicle dynamics into mathematical formulations, often used to enable overtaking maneuvers.  Several control approaches have been proposed for overtaking maneuvers. Model Predictive Control (MPC) strategies are discussed in \cite{r19}-\cite{r21} to handle trajectory planning and obstacle avoidance. Other methods include stochastic MPC \cite{r22}, sliding mode control \cite{r24}, and Mixed-Integer Quadratic Programming \cite{r25}. A comprehensive survey is provided in \cite{r27}. 
%\marginnote{\textcolor[rgb]{1,0,0}{[\textbf{R1-1}], [\textbf{R3-1}], \textbf{[R3-3]} }}
%\textcolor[rgb]{1,0,0}{While current AI-based and mathematical control methods offer valuable solutions for autonomous overtaking, they frequently face challenges in simultaneously ensuring real-time computational efficiency and robust constraint satisfaction under dynamic uncertainties.}
In recent years, hierarchical and integrated Deep Reinforcement Learning (DRL) frameworks \cite{[R1], [R2]} have been proposed to manage overtaking sub-tasks and enforce physical constraints. However, despite their excellent reactivity, these data-driven approaches frequently fail to provide the rigorous mathematical guarantees on state and input constraints provided by formal model-based methods.
%Recently, a Dynamic Option Policy enabled Hierarchical Deep Reinforcement Learning (DOP-HDRL) framework was proposed in \cite{[R1]} to decompose the overtaking maneuver into simpler sub-tasks managed by dedicated RL agents. Similarly, an integrated RL framework is introduced in \cite{[R2]}, featuring dynamic state mapping for raw track observations and an action mapping technique to strictly enforce physical traction constraints while blending conservative controllers with RL policies to guide safe exploration. However, despite their excellent reactivity, such data-driven approaches still struggle to inherently provide the rigorous mathematical guarantees on state and input constraints ensured by formal model-based control methods.
Within the MPC domain, recent overtaking algorithms \cite{[R3], [R4]} offer valuable solutions. However, to bypass non-convexity and maintain real-time feasibility, they often require significant simplifications such as decoupled point-mass models \cite{[R3]} or complex switching heuristics \cite{[R4]}.
%Within the Model Predictive Control domain, a recent trajectory planning algorithm for overtaking was introduced in \cite{[R3]}. To mitigate computational burdens, this method decouples the problem into loosely coupled longitudinal and lateral MPCs, by using a simplified point mass model and linear collision avoidance constraints. Furthermore, to streamline decision-making during autonomous overtaking, an MPC-based switching control framework is proposed in \cite{[R4]} as an alternative to conventional rule-based methods, bridging high-level tactical decisions with low-level control inputs.
%While these recent MPC strategies offer valuable solutions, they often require significant model simplifications or complex switching heuristics to bypass non-convexity and maintain real-time feasibility. 
Starting from this analysis, we focus on constrained receding horizon control approaches that are particularly attractive for obstacle avoidance in motion planning because they inherently generate feasible trajectories at every time step, ensuring safe progress toward a specified goal \cite{r28,r32}-\cite{r32}. %},r30,r31,r32}. 
The main contribution of this paper lies in overcoming the traditional limitations of the Set-Theoretic Fast Ellipsoidal MPC (ST-FE-MPC). Inspired by \cite{angeli1} -which was strictly limited to stabilization- this work significantly extends the theory to address reference tracking and motion planning for autonomous vehicles modeled as uncertain polytopic systems. Unlike standard optimization-based approaches, our architecture transforms the online control phase into a highly efficient set-membership evaluation against a structured sequence of robust invariant ellipsoidal sets. This inherently guarantees recursive feasibility, strict state and input constraint satisfaction, and finite-time maneuver completion, effectively solving the complex obstacle avoidance problem without the need for computationally heavy online non-convex optimization.

The proposed control architecture was validated through a dual-stage evaluation process. Initial testing was performed using the MATLAB Automated Driving Toolbox (ADT) in Simulink to assess core functionality. Subsequently, high-fidelity co-simulations in the CARLA 3D environment were conducted to rigorously test the controller against complex physics, unmodeled dynamics, and dynamic traffic conditions. This comprehensive validation confirms the controller’s robustness and its suitability for real-world autonomous driving applications.

The remainder of this paper is organized as follows. Section \ref{s1} introduces the necessary preliminaries and notation. In Section \ref{s2}, the problem is formally stated, while Section \ref{setheo} presents the control procedure. The set-theoretic fast ellipsoidal MPC approach of \cite{angeli1} is here briefly recalled and subsequently extended to obstacle avoidance and motion planning. Section \ref{s3} describes the modeling approach, specifically presenting a suitable state-space representation and the corresponding uncertain polytopic approximation of the vehicle's lateral and longitudinal dynamics. Finally, Section \ref{s6} reports simulation results and comparisons with other MPC strategies to demonstrate the effectiveness and advantages of the proposed control scheme.
%In particular, the proposed control architecture was validated using the MATLAB Automated Driving Toolbox (ADT) within the Simulink environment, which enables the simulation of realistic driving scenarios. Furthermore, high-fidelity co-simulations were conducted in the CARLA 3D environment to rigorously test the controller's performance against complex physics, unmodeled dynamics, and dynamic traffic conditions, ensuring its practical viability for autonomous driving. 
Finally, some conclusions end the paper.
%%%%%%%%%%%%%%%%%%%
\section{Preliminaries and Notations}
\label{s1}
This section presents a series of notations useful to understand the key aspects of the control design reported throughout the remainder of the paper.
\subsection{Uncertain Polytopic Vehicle's Model Approximation}
We will focus on autonomous driving vehicles modeled as uncertain polytopic discrete-time linear systems of the 
form
\begin{equation}
		x(t+1)=\Phi\left( p\right)x(t)+G\left(p \right)u(t)+G_d\left(p \right)d(t)
		\label{Sp1}
	\end{equation}
where $t \in \mathbb{Z}_+:=\left\lbrace0,1,... \right\rbrace$, $x(t)\in\mathbb{R}^n$ is the state, $u(t)\in\mathbb{R}^m$ the control input, $d(t)\in\mathbb{R}^{n_d}$ an exogenous disturbance and $p\in\mathbb{R}^l$ a possibly time-varying scheduling parameter. We assume that  
\begin{equation}\label{d}
d(t)\in\mathcal{D}\subset\mathbb{R}^{n_d}, \forall t \in \mathbb{Z}_+ 
\end{equation}
represents persistence disturbances with $\mathcal{D}$ compact with $ 0_n\in\mathcal{D}$ and the system matrices $\Phi\left( p \right)$, $G\left(p\right)$ and $G_d\left(p\right)$ belong to the polytopic matrix family
\small
	\begin{equation}
		\Sigma\left( \mathcal{P}\right) :=
		\left\lbrace 
		\left(\Phi(p),G(p),G_d(p)\right)=
		\overunderset{k}{j=1}{\sum}p_j
		\left(\Phi_j,G_j,G_{dj}\right),p\in\mathcal{P} 
		\right\rbrace 
	\end{equation}
	\normalsize
with the scheduling vector $p=\left[p_1,\ldots,p_k\right]^T$  belonging to the unit simplex
	\begin{equation}
		\mathcal{P}:=\left\lbrace p\in\mathbb{R}^k: 0 \le p_j \le 1,\,\,\overunderset{k}{j=1}{\sum}p_j=1\right\rbrace. 
	\end{equation}
It also assumed that the system is subject to input and state set-membership constraints
\begin{eqnarray}
	u(t) &\in& \mathcal{U}\subset\mathbb{R}^{n}, \quad \forall t \geq 0,\\
	x(t) &\in& \mathcal{X}\subset\mathbb{R}^{m}, \quad \forall t \geq 0,\label{eq:ux_constr}
\end{eqnarray}
with $\mathcal{U}$ and $\mathcal{X}$ convex and compact sets containing their zero vectors.
	
% DEFINITION OBSTACLE SCENARIO
%\begin{definition}
\begin{figure}[t]
	\centering
		\includegraphics[width=0.5\textwidth]{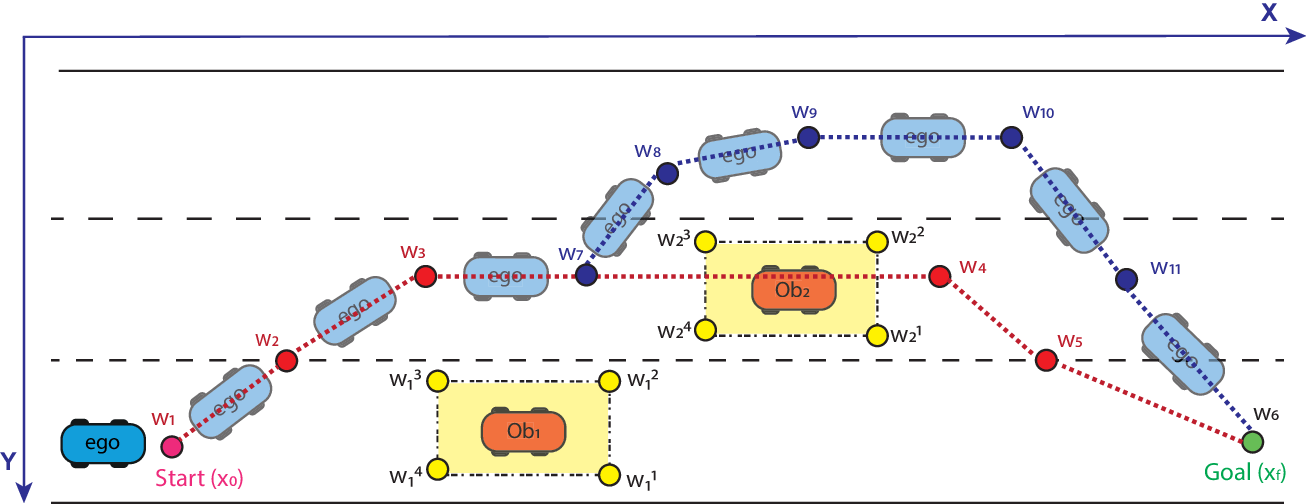}
	\caption{Way-points ($\mathcal{W} = \left[w_1,\ldots,w_6\right]$ (red trajectory) describing a feasible paths; way-points $\mathcal{W}' = \left[w_7,\ldots,w_6\right]$ (blue trajectory) describing an alternative feasible paths; points ($\mathcal{W}_{1} = \left[w_{1}^1,\ldots,w_{1}^4\right]$; $\mathcal{W}_{2} = \left[w_{2}^1,\ldots,w_{2}^4\right]$) describing the obstacle scenario $\mathcal{O}=\left\{Ob_1,Ob_2\right\}$.}%; way-points ($\mathcal{W}_{sw} = \left[w_{sw_1},\ldots,w_{sw_2}\right]$) referred to a switching sequence.}
	\label{fig:waypoints}
\end{figure}
\subsection{Obstacles}
As depicted in Figure \ref{fig:waypoints} where a normal driving situation is reported, lead vehicles traveling at lower speeds in the same lane can be treated as dynamic obstacles that the ego vehicle must overcome. Formally, let $\mathcal{O}$ denote a specific obstacle scenario represented by a collection of planar bounding boxes. Relying on real-time radar measurements, the ego vehicle's adaptive cruise control (ACC) adjusts its longitudinal speed to maintain a predefined safe distance from these obstacles, or otherwise proceeds at a constant reference speed.
%In a normal driving situation, like the one shown in Fig. \ref{fig:waypoints}, vehicles  in the same lane moving at lower speeds (lead cars) may be seen as obstacles that the present vehicle must pass over. Thus, in order to be more rigorous, let $\mathcal{O}$ denote a specific obstacle scenario consisting of a collection of different  physical obstacles represented by planar yellow boxes as the ones depicted in Fig. \ref{fig:waypoints}. Arguably, these obstacles are leading vehicles proceeding on the same lane that are detected in real-time by the ego car via a radar sensor that provides measurements of the relative distance between the ego vehicle and the lead vehicles. According to these measurements, the ego vehicle, operated by an adaptive cruise control system, decreases its speed when the relative distance to the preceding vehicle is less than or equal to a specified safe distance. Otherwise, the ego car proceeds at a constant reference speed.

In order to obtain an obstacle representation using directly measurable quantities, it is possible to define the avoidance conditions in terms of the relative longitudinal  \( x^{rel}(t) = X^{lead}(t)-X^{ego}(t) \) and lateral  \( y^{rel}(t) = Y^{lead}(t)-Y^{ego}(t) \) distances. Specifically, \( x^{rel} \) denotes the relative distance between the ego vehicle and the detected object along the longitudinal axis, and $y^{rel}$ along the lateral axis. 

To facilitate obstacle avoidance in a modular and scalable manner, it is feasible to adopt a formulation that addresses obstacles individually - one at a time - utilizing sensor input, including radar, LiDAR, and camera. While the coordinates \( X^{ego}(t) \), \( Y^{ego}(t) \) are obtained in real time from the vehicle’s GPS system, the coordinates of the obstacle center \( (x^{rel}, y^{rel}) \) are assumed to be available from onboard sensors. Based on this, we define the following region:
\begin{equation}
\mathcal{O} :=
\left\{
\begin{bmatrix}
\delta x \\
\delta y
\end{bmatrix} \in \mathbb{R}^2 :
| \delta x - x^{rel} | \leq \delta_x,\quad
| \delta y - y^{rel} | \leq \delta_y
\right\}
\label{eq:def1o}
\end{equation}
in terms of the slack variables $(\delta x, \delta y)$ to characterize the obstacles where $\delta_x$ and $\delta_y$ represent the required safety margins. To ensure strict collision avoidance in real-world autonomous driving, these margins are not merely pragmatic bounding boxes, but are rigorously defined to account for both the physical vehicle kinematics and the inherent measurement noise of the perception system. Specifically, the longitudinal margin can be formulated as 
$$\delta_x = \frac{L_{ego} + L_{lead}}{2} + \epsilon_{x},$$
 where $L_{ego}$ and $L_{lead}$ are the physical lengths of the ego and leading vehicles, and $\epsilon_{x}$ represents the maximum bounded uncertainty derived from the sensor's error model (e.g., a $3\sigma$ confidence interval for radar/LiDAR spatial resolution and bounding-box estimation noise). A similar formulation applies to the lateral margin $\delta_y$ based on the vehicle widths and lateral sensor variance. This ensures that the robust invariant sets inherently absorb worst-case sensor inaccuracies.

Notice that the ego car and the obstacle are moving at the same speed when $(x^{rel} , y^{rel})$ remain constant during the motion. 

Accordingly, the obstacle-free region can be defined as:
\begin{equation}
\mathcal{O}^{\text{free}} :=
\left\{
\begin{bmatrix}
\delta x \\
\delta y
\end{bmatrix} \in \mathbb{R}^2 :
| \delta x - x^{rel} | > \delta_x,\quad
| \delta y - y^{rel} | > \delta_y
\right\}
\label{eq:def2o}
\end{equation}

Then, by assuming that each obstacle is evaluated individually based on its relative coordinates \( (x^{rel}, y^{rel}) \), we also may argue that these components could be present in the state vector \( x \in \mathbb{R}^n \) of the ego car. Then,  a selection matrix \( P \in \mathbb{R}^{2 \times n} \), could be introduced that maps $x$ into its components \( (x^{rel}, y^{rel}) \). Then, the safe set can be expressed directly in terms of the system's state as 
\begin{equation}
\mathcal{X}^{\text{free}} :=
\left\{ x \in \mathbb{R}^n : P x \notin \mathcal{O} \right\}
\label{eq:def3oe}
\end{equation}
%\hl{where} \( H_i \in \mathbb{R}^{4 \times 2} \) and \( b_i \in \mathbb{R}^4 \) define the bounding box around the $i$-th obstacle, typically %using the same structure
%\[
%H_i =
%\begin{bmatrix}
%\phantom{-}1 & \phantom{-}0 \\
%-1 & \phantom{-}0 \\
%\phantom{-}0 & \phantom{-}1 \\
%\phantom{-}0 & -1
%\end{bmatrix},\quad
%b_i =
%\begin{bmatrix}
%-\delta_x \\
%-\delta_x \\
%-\delta_y \\
%-\delta_y
%\end{bmatrix}
%\]
%Notice that each obstacle \( i \) defines a local unsafe region \( \{ x : H_i B_{rel} x \leq b_i \} \), and the global safe set is obtained by %excluding all of them simultaneously.
\\
\textit{Remark 1 - Note that, each obstacle is conservatively modeled as a rectangular region aligned with the coordinate axes. Since such sets are already convex by construction, no additional convexification procedure is required.
}

%\begin{equation}
%		 \mathcal{O}^o_{free}:=\left\lbrace (X,Y) \in \mathbb{R}^n:h_o(X,Y)>0 \right\rbrace.
%		\label{eq:def2}
%\end{equation}
%\end{definition}

% DEFINITION ROBUSTLY INVARIANT SET
%\begin{definition}
\subsection{Robust Control Invariant Sets}
A set $\mathcal{T}$ is said to be \textit{robustly control invariant} with respect to the uncertain polytopic system
\begin{equation}
	x(t+1)=\Phi(\mathcal{P})x(t)+G(\mathcal{P})u(t) + G_d(\mathcal{P})d(t),
\label{eq:def0}
\end{equation}
if, for all $x \in \mathcal{T}$ and for any disturbance  $d \in \mathcal{D}$ exits a $u(t)$ such that 
\begin{equation}
	\Phi(\mathcal{P})x(t)+G(\mathcal{P})u(t) + G_d(\mathcal{P})d(t) \subset \mathcal{T},
\label{eq:def1}
\end{equation}
This ensures that, under any admissible disturbance $d(t)$ the state evolution remains within $\mathcal{T}$, preserving its invariance.
Then, given a robustly control-invariant region $\mathcal{T}$, we define the sets of states that can be driven into $\mathcal{T}$ within $i$ steps, regardless of disturbances and uncertainties acting on the system, through the following recursive formulation
\begin{eqnarray}	\label{recursion}
%	\begin{array}{lcl}
		\mathcal{T}_0 & := &\mathcal{T} \nonumber \\
		\mathcal{T}_i & := & \left\{ x \in \mathcal{X}: \exists u \in \mathcal{U}: \forall d \in \mathcal{D},\right. \nonumber \\
		& & \left.\forall p \in \mathcal{P}, \Phi(p) x +G(p)u+G_d(p)d \subset \mathcal{T}_{i-1} \right\} 
	%							\end{array}
\end{eqnarray}
where:
\begin{itemize}
	\item  $\mathcal{T}_0$ is known as the terminal region; 
	\item $\mathcal{T}_i$  is the set of states that can be driven into $\mathcal{T}_{i-1}$, in a single control step, regardless of disturbances and parametric uncertainties, while satisfying causality constraints. By induction, it follows that $\mathcal{T}_i$ characterizes the set of states that can be driven into $\mathcal{T}$ in at most $i$ control steps, with  $\mathcal{T}_i \subset \mathcal{T}_{i+1}$ \cite{angeli1}. Moreover, if $ \mathcal{X}$ is closed and bounded $\mathcal{T}_\infty:=\lim_{i\to\infty}\mathcal{T}_i$ exists.
\end{itemize}
Because the shapes of $\mathcal{T}$ grows in complexity with $i$, one of the main results in \cite{angeli1} was the determination of a variant of recursions (\ref{recursion}) that employees maximum volume ellipsoidal inner approximations $\mathcal{E}_i$, allowing a constant number of parameters at each iteration for their characterization:
\begin{eqnarray}\label{recursion2}
\mathcal{E}_0 & := &\mathcal{E} \nonumber \\
\mathcal{E}_i & :=& In\left[ \left\{ x \in \mathcal{X}: \exists u \in \mathcal{U}: \,\,\forall d \in \mathcal{D} \right.\right., \nonumber \\
 && \left.\left. \forall p \in \mathcal{P} \Phi(p) x +G(p)u+G_d(p)d \subset \mathcal{E}_{i-1} \right\}\right] \nonumber \\
			&=& x^TP^{-1}_{i}x\leq 1	
\end{eqnarray}
with $\mathcal{E}_{i-1}\subset \mathcal{E}_{i}\subset \mathcal{E}_{\infty}$,  where $In\left[\cdot\right]$ denotes the inner ellipsoidal approximation and $P_i=P_i^T\geq 0$ a sequence of symmetrical positive defined shaping matrices that characterize the ellipsoidal sequence. See \cite{angeli1} for construction details.

\subsection{Set-Theoretical Fast Ellipsoid MPC approach}
\label{setheo}
Here, the ST-FE-MPC approach of \cite{angeli1} is briefly recalled. According to the uncertain polytopic system description (\ref{Sp1}). 
it is required to solve a $0_x$-stabilization problem under the input and state constraints (\ref{eq:ux_constr}).
A way to address the above control  problem is to consider a dual-mode MPC approach. Basically, a final robust control-invariant set $\mathcal{T}$ is introduced, where the system trajectories are ultimately trapped under a chosen  control law, e.g. $u(t)=Kx(t)$. Moreover,  given $\mathcal{T}$, we could compute the sets of states controllable in $i$ steps to $\mathcal{T}$.
The developments of previous  subsection allow one to synthesize very easily an on-line Receding-Horizon control strategy. Specifically, at each time $t$, the smallest index $i$ such that $x(t) \in \mathcal{E}_{i}$  is first determined. Then, a command $u(t)$ is determined by minimizing a convenient running cost. Notice that the determination of the ellipsoidal sets $\mathcal{E}_{i}$ does not depend on the running cost to be used during the on-line operations. This allows more flexibility in the choice of the cost which could eventually even be changed on-line along the system trajectories. Then, the corresponding online procedure is as follows:
\noindent \rule{\columnwidth}{0.05cm}
\noindent \textbf{RHC Algorithm - Online Phase}\\
\noindent \rule{\columnwidth}{0.05cm}
\begin{enumerate}\label{alg}
\item[(1)] $t=0$
\item[(2)] Let $i(t)=\min\{i: x(t)\in \mathcal{E}_{i}\}$
\item[(3)] If $i(t)=0$ then $u(t)=Kx(t)$
\item[(4)] Else
\begin{description}
\item $u(t)= \mbox{argmin}_{u} J_{i(t)}(x(t), u)$
\item \hspace{1cm} subject to
\item $\Phi_j x(t) + G_j u \in \mbox{In} \left[ \mathcal{E}_{i-1} \sim G_{d_i} \mathcal{D} \right] ,\,\,j=1,2,...., l, \,\, u\in \mathcal{U}$
\end{description}
\item[(5)] apply $u(t)$; $t=t+1$; goto 2
\end{enumerate}
\noindent \rule{\columnwidth}{0.05cm}
\\
In the previous algorithm, $\mbox{In} \left[ \mathcal{A} \sim \mathcal{B} \right]$ denotes a suitable inner ellipsoidal approximation  of the set $\mathcal{A} \sim \mathcal{B}$ and the operator $\sim$ denotes subtraction between set ($\mathcal{A} \sim \mathcal{B}=\{ a: a+b\in \mathcal{A}, \forall b\in \mathcal{B}\})$ \cite{Kur}. For a discussion on
the selection of $J_{i(t)}(x(t), u)$  see  \cite{angeli1}, where several choices were discussed.

\section{Problem Statement and Control Design}
\label{s2}
Building upon the OAMP framework introduced in \cite{OAMP} for mobile robots modeled as LTI systems, this section extends the approach to autonomous vehicles modeled as uncertain polytopic systems executing overtaking maneuvers. Accordingly, the problem is formally stated as follows.\\
%Obstacle avoidance and motion planning (OAMP) problem is a key challenge in autonomous driving vehicle, enabling safe navigation through detection and interpretation of obstacles. Overtaking maneuvers build on these capabilities, requiring precise control to pass slower vehicles. Here, the solution to the OAMP problem presented in \cite{OAMP} for mobile robots modeled as LTI systems, is outlined for autonomous driving vehicle modeled as uncertain polytopic systems. In this respect, the following problem can be stated. \\
%
%\begin{definition} 

\noindent \textbf{OAMP Problem} - \textit{With reference to Figure \ref{fig:overtaking}, consider an autonomous (ego) vehicle driving in traffic context. Given the polytopic family of discrete-time linear systems (\ref{Sp1}) and set of obstacle (lead vehicle) scenarios (\ref{eq:def1})}.
%Assume that each $Ob_j^i$ has a polyhedral convex structure described as the intersection of $l_j$ half-spaces:
%\begin{equation}
%	Ob_j^i:
%	\left[\begin{array}{c}
%		\left(H_j^i\right)^T_1 \\ \vdots \\ \left(H_j^i\right)^T_{l_j}
%	\end{array}\right] z \leq 
%	\left[\begin{array}{c}
%		\left(g_j^i\right)^T_1 \\ \vdots \\ \left(g_j^i\right)^T_{l_j}
%	\end{array}\right]
%\label{eq:halfspaces}
%\end{equation}
%where $z: = Bx \in \mathbb{R}^2$ are the planar components of the state space $x \in \mathbb{R}^n$ and $B \in \mathbb{R}^{2\times n}$ a projection matrix.
\textit{Determine a state-feedback control law}
\begin{equation}
\begin{array}{l}
	u(t) = g(x(t)) \\[8pt]
	\quad s.t. \\[8pt]
\end{array}
\label{eq:control_law}
\end{equation}
\begin{equation}
\begin{array}{l}
	u(t) \in \mathcal{U}, \quad \forall t \geq 0,\\
	x(t)\in \mathcal{X}, \quad \forall t \geq 0,
\end{array}
\label{eq:ux_constr1}
\end{equation}
\textit{with} $\mathcal{U}$ \textit{and} $\mathcal{X}$ \textit{compact subsets of} $\mathbb{R}^m$ \textit{and} $\mathbb{R}^n$, \textit{respectively, such that starting from an initial condition} $x_0$ \textit{the vehicle trajectory} $x(t)$ \textit{is driven to a target position} $x_f$ \textit{regardless of any obstacle scenario occurrence}.\\

To solve the OAMP problem, we propose a Set-Theoretic Fast Ellipsoidal MPC (ST-FE-MPC) strategy for autonomous vehicles modeled as uncertain polytopic systems. By leveraging sequences of inner ellipsoidal approximations of one-step controllable sets, this approach guarantees robust constraint satisfaction and computational tractability against disturbances and dynamic obstacles.
%To effectively address the challenges posed by OAMP Problem, we propose a control strategy based on the Set-Theoretic Fast Ellipsoidal Model Predictive Control (ST-FE-MPC) framework. This approach is specifically designed for autonomous vehicles modeled as uncertain polytopic systems, ensuring robust constraint satisfaction while maintaining computational tractability. By leveraging sequences of inner ellipsoidal approximations of one-step controllable sets, the proposed method systematically accounts for state and input constraints, disturbances, and dynamic obstacle scenarios that may arise during overtaking maneuvers. In what follows, an application ST-FE-MPC strategy ensuring safe and feasible trajectory planning under time-varying obstacle (traffic) conditions is presented.

\subsection{Computation of feasible trajectories}
Figure \ref{fig:waypoints} illustrates the feasible trajectory computation via a dynamic overtaking scenario. The ego vehicle, initially at $x_0$, targets $x_f$ but encounters obstacle $Ob_1$. It computes an initial overtaking path ($w_1 \to w_2 \to w_3$) into the center lane. When a new dynamic obstacle $Ob_2$ appears, invalidating the planned route, the vehicle dynamically re-plans a secondary maneuver into the leftmost lane ($w_7 \to \dots \to w_{11}$), safely reaching the final position $w_6 = x_f$.

Based on this context, and given the initial and final states $x_0$ and $x_f$ and the obstacles scenarios, the objective is to design a procedure that:
\begin{enumerate}
	%\item Generates intermediate path segments around the obstacle with the aim of encapsulate it inside specified sequence;
	\item Generates intermediate path segments between consecutive way-points, ensuring local feasibility and satisfaction of system constraints;
	\item Recursively composes these segments to construct a globally feasible trajectory that connects the initial and final configurations while avoiding obstacles.
\end{enumerate}
To this end, we assume that a set of way-points - $\mathcal{W} = [w_1, \ldots, w_\ell]$, with $w_1 = x_0$ as the initial point and $w_\ell = x_f$ as the final goal, defining a candidate path (or sub-path) to be followed - is available.

Given a set of way-points $\mathcal{W}$, the proposed procedure constructs a feasible trajectory by propagating backward from the goal $x_f$ to the initial state $x_0$. By exploiting an extended set-theoretic robustly invariant framework, it progressively defines intermediate safe regions around equilibrium points, inherently ensuring feasible transitions despite dynamic obstacles and disturbances.
%On the basis of previous discussion, the proposed procedure constructs a feasible trajectory by propagating backward from the goal state ($x_f$) toward the initial state ($x_0$), progressively defining intermediate safe regions centred on different equilibrium points. At each step, the method ensures that transitions between successive regions are feasible, even in the presence of dynamic obstacles or external disturbances.
%Then, given a set of way-points ($\mathcal{W}$, etc.), and any pair of feasible initial and final state vectors belonging to these sets, the obstacle avoidance and motion planning control problem can be addressed using an extended formulation of the set-theoretic robustly invariant framework introduced in the previous section.
Specifically, for each given point, it is possible to construct a sequence of ellipsoidal inner approximations of the $i$-step ahead \textit{robustly control invariant obstacle-free} ellipsoids, recursively defined as:
\normalsize
\begin{eqnarray}\label{rec}
\mathcal{E}^s_0 & \! \! \! := \! \! \! &\mathcal{E} \nonumber\\
\mathcal{E}^s_i& \! \! \! := \! \! \! & \left\{ x-x^s_{eq} \in \mathcal{X} \cap \mathcal{X}^{\text{free}} : \exists (u -u^s_{eq})\in \mathcal{U}, \forall d \in \mathcal{D},\right. \nonumber \\ 
&& \left. \forall p \in \mathcal{P}, \Phi(p)(x-x^s_{eq}) + G(p)(u-u^s_{eq}) +\right. \nonumber \\ 
&&\left.G_d(p)d \subset \mathcal{E}^s_{i-1} \right\} \nonumber \\
						& \! \! \! = \! \! \!& (x-x^s_{eq})^TP^{-1}_i(x-x^s_{eq}) \leq 1 
\end{eqnarray}
\normalsize
where $(x^s_{eq},u^s_{eq})$ defines a suitable equilibrium point over which the ellipsoids are centered. This recursive formulation, if non-empty, allows the construction of a family of $x^s_{eq}$-centred ellipsoidal sets that are by construction each one recursively one-step ahead controllable and robustly invariant with respect to  all admissible disturbances $d \in \mathcal{D}$ and parameter variations $p \in \mathcal{P}$ while ensuring constraint satisfaction and obstacle avoidance at every step.
%\begin{definition}
%In this respect, the following \textbf{Algorithm 1} can be seen as a useful tool to determine an intermediate path between two given points.
Intuitively, the trajectory generation process can be viewed as a multi-tier architecture involving three distinct layers: Algorithm 1 serves as a local planner, computing a 'safe virtual tube' of nested invariant ellipsoids to connect adjacent spatial points; Algorithm 2 acts  as a global planner, iteratively stitching these local tubes to form a continuous, collision-free corridor from start to goal; and Algorithm 3 operates as a real-time navigator, dynamically steering the vehicle's state through this predefined corridor while rejecting local disturbances.
%From an intuitive perspective, the overall trajectory generation can be viewed as a multi-tier process involving three distinct layers: \textbf{Algorithm 1} that acts as a local planner, computing a "safe virtual tube" of nested invariant ellipsoids that connects two adjacent points in space. \textbf{Algorithm 2} acting as the global planner, iteratively stitching these local tubes together along the sequence of predefined waypoints to form a continuous collision-free corridor from the start to the goal. \textbf{Algorithm 3} that operates online as a real-time navigator, dynamically pushing the vehicle's state through this predefined safe corridor while rejecting local disturbances.}
%
\begin{figure}[htbp]
	\centering
		\includegraphics[width=0.50\textwidth]{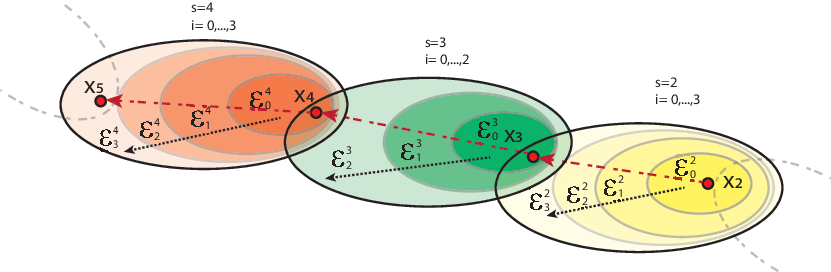}
	\caption{Sequence of ellipsoidal inner approximations construction.}
	\label{fig:ellipsoid}
\end{figure}

More in details, by using recursion (\ref{rec}),  \textbf{Algorithm 1} allows the determination of: 
\begin{itemize}
	\item the sequence ellipsoidal family $\left\{\mathcal{E}_i^s\right\}$;
	\item the sequence of equilibrium points $\left\{x_{eq}^s\right\}$; %where $\left\{\mathcal{E}_i\right\}$ are centered;
	\item the sequence of stabilizing feedback gains  $\left\{K^s\right\}$. %such that:
		%\begin{equation}
		%	(\Phi(p)+G(p)K_i)\mathcal{E}_i^0+G_d(p)\mathcal{D} \subset \mathcal{E}_{i-1}^.
		%\label{eq:closed}
		%\end{equation}
\end{itemize}

Figure~\ref{fig:ellipsoid} provides a graphical representation of the construction of a sequence of ellipsoidal inner approximations. Specifically, the figure illustrates a family of ellipsoidal sets, each identified by two indices:
\begin{itemize}
	\item $s$, denoting the index of the ellipsoidal family;
	\item $i$, indicating the specific ellipsoid within the $s$-th family.
\end{itemize}
For instance, when $s = 4$, the procedure generates a sequence $\left\{\mathcal{E}_i^{s=4}\right\}$ centered at $x_4$ and iteratively expands the ellipsoidal regions until the point $x_3$ is included. Next, a new sequence $\left\{\mathcal{E}_i^{s=3}\right\}$ is generated, centered at $x_3$, and the process is repeated until the subsequent point is reached. This backward propagation continues through successive families until the initial point $x_1$ is encompassed.

\noindent \rule{\columnwidth}{0.05cm}
\noindent \textbf{Algorithm 1} - One-step sequence computation\\
\noindent \textbf{Input}: $x_{1}$, $x_2$, $s_{start}$;\\
\noindent \textbf{Output}: $\left\{x_{eq}^s\right\}$, $\left\{K^s\right\}$, $\left\{\mathcal{E}_i^s\right\}$, $N_s$\\
\noindent \rule{\columnwidth}{0.05cm}
\begin{enumerate}
	\item[(1)] Given $x_1$ and $x_2$; 
	\item[(2)] Design the set $\mathcal{E}_0^0$ centered at $x_2$ and the stabilizing gain $K^0$ satisfying constraints (\ref{eq:ux_constr1});
	\item[(3)] Set $s = s_{start}$ and $x_{eq}^s=x_2$;
	%\item[(4)] Store $s = 0$ in the index vector $IR^s$;
	\item[(4)] \textbf{While} $x_1 \notin \mathcal{E}_s^{N_\mathcal{E}^s}$    
		\begin{itemize}
			\item[-] Derive the sequence $\left\{\mathcal{E}_i^s\right\}_{i=1}^{N_\mathcal{E}^s}$ with initial value $\mathcal{E}_{0}^s$ using recursion (\ref{rec});
			\item[-] $x_{eq}^{s+1} :=\underset{x \in \mathcal{E}^{s}_{N_\mathcal{E}^s}}{\mathrm{argmin}} \left\|x - x_1\right\|_2 $; 
			\begin{itemize}
				\item [] \textbf{If} $ x_{eq}^{s+1} == x_{eq}^{s}$ then \textbf{stop}
			\end{itemize}
			\item[-] Compute new pair $(K^{s+1}, \mathcal{E}^{s+1}_0)$ with $\mathcal{E}^{s+1}_0$ the ellipsoidal inner approximations centered at $x_{eq}^{s+1}$;
			%\item[-] $\mathcal{E}_{i+1}^0 = \tilde{\mathcal{E}}$;% and $m\leftarrow m+1$;
			\item[-] $s\leftarrow s+1$;
			%\item[-] $s_{start}=s$;
			%\item[-] Store $\left\{ 1,\ldots, N_s \right\}$ into $IR^s$;
 			%\item[] Set $x_f = x(0)$;
			%\item[] Set $x_{eq}^m=x_{eq}^{m+1}$;
		\end{itemize}
		%\item[(6)] Store $m = 0$ in index vector $IR_i$.
		\item[(5)]\textbf{EndWhile}
		\item[(5)]Set $N_s= s$.
\end{enumerate}
\noindent \rule{\columnwidth}{0.05cm}
\\
Note that, in \textbf{Algorithm 1}: 
\begin{itemize}
	\item $N_\mathcal{E}^s$ is an integer representing the saturation level for the region growth of the $s$ family. 
	%\item the vector $IR^s$ has the aim to keep trace of all computed robust positively invariant regions;
	\item the pair $(K^{s}, \mathcal{E}^{s}_0)$ are compute via standard LMI techniques \cite{kothare}.
\end{itemize}

Then, to construct a globally feasible trajectory that connects the initial and final configurations while avoiding obstacles it is possible to account the procedure described by \textbf{Algorithm 2}. This algorithm recursively exploits \textbf{Algorithm 1} to the determine the full path by accounting the set of points representing the obstacles, the transitions and and the trajectory to be followed.

%Note that, in what follows an additional vector $\left\{IR_i^s\right\}$  is introduced, that stores $IR_i$ for all obstacles scenario $\mathcal{O}^o$,

\noindent \rule{\columnwidth}{0.05cm}
\noindent \textbf{Algorithm 2 - Full path (FP) computation}\\
\noindent \textbf{Input}: $\mathcal{W}_p$;\\
\noindent \textbf{Output}: $FP$\\
\noindent \rule{\columnwidth}{0.05cm}
\begin{enumerate}
	\item[(1)] Given the set of points $\mathcal{W} = \left[w_{1},\ldots,w_{\ell}\right]$. \\
		%Note that, $\mathcal{W}_p$ refers, alternatively, to $\mathcal{W}$, $\mathcal{W}_{obj}$, or $\mathcal{W}_{sw}$;% and the set of points $\mathcal{W}^p = \left[w_1^p,\ldots,w_\ell^p\right]$ representing, respectively, the path to be followed and the vertex of the polytope containing the obstacle;
	\item[(2)] Define the full path as the index vector FP;
	\item[(3)] Set $N_s= 0$;
	\item[(4)] \textbf{For} $wp=\ell-1:1$   
		\begin{itemize}
			\item[(a)] Set $x_1=w_{wp}$, $x_2=w_{wp+1}$ and $s_{start} = N_s$;
			\item[(b)] Compute $\left\{x_{eq}^s\right\}$, $\left\{\mathcal{E}_i^s\right\}$, $\left\{K^s\right\}$ and $N_s$ using \textbf{Algorithm 1};
			\item [(c)] $FP(wp) =\left(\left\{x_{eq}^s\right\}, \left\{\mathcal{E}_i^s\right\}, \left\{K^s\right\}\right)$;
		\end{itemize}
		\item[(5)]\textbf{EndFor}
\end{enumerate}
\noindent \rule{\columnwidth}{0.05cm}
\\
Note that, at the last iteration of Algorithm 2 $N_s$ stores the total number of intermediate paths;

\begin{figure}[t]
	\centering
		\includegraphics[width=0.5\textwidth]{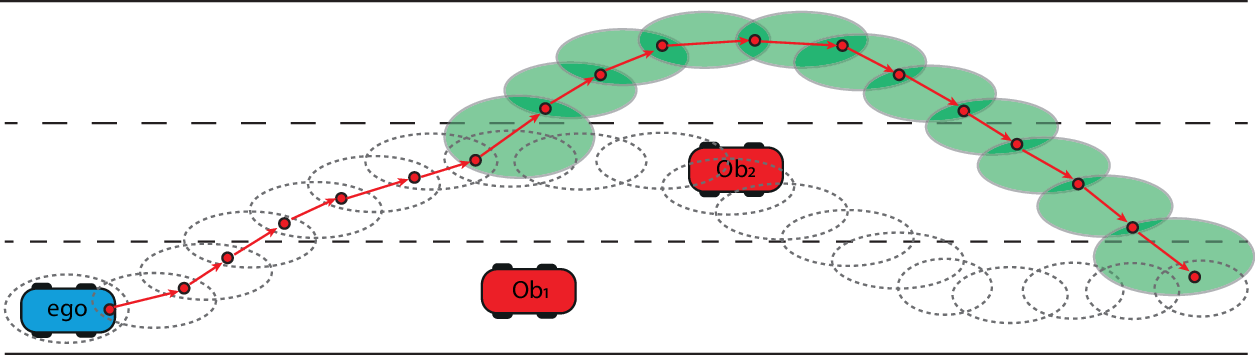}
	\caption{Full path generation: ellipsoid set $\mathcal{E}_i^s$ sequence (white ellipsoid) generated in response to the presence of the vehicle $Ob_1$, updated ellipsoid set $\mathcal{E}_i^s$ sequence (green ellipsoid) accounting the new vehicle $Ob_2$, final path (red arrows). }
	\label{fig:scenario_change2}
\end{figure}
%% DA QUI IN POI DESCRVERE A PAROLE E CON RIFERIMETNO ALLE FIGURE LA COSTRUZIONE DELLA SEQUENZA DI OSTACOLO E DELLA SEQUNZA DI SWITCHING

It is important to note that, the procedure described by \textbf{Algorithm 1} and \textbf{2} can be viewed as a generic procedure that allows one to determine the ellipsoidal family and stabilizing feedback gains given any generic set of points. As an example, with reference to Figure \ref{fig:waypoints} the procedure allows one to:
\begin{itemize}
	\item determine the ellipsoid sequence \(\left\{\mathcal{E}_i^{s}\right\}\) pertain to the way points $\mathcal{W} = \left[w_1,\ldots,w_6\right]$ (red trajectory) that accounts for the presence of the obstacle \(Ob_1\);
	\item update the ellipsoid sequence \(\left\{\mathcal{E}_i^{s}\right\}\) by considering the way points $\mathcal{W}' = \left[w_7,\ldots,w_6\right]$ (blue trajectory) that accounts for the presence of the new obstacle \(Ob_2\);
	%\item the obstacle sequences $\left\{\mathcal{E}_i^{O^{s,o}}\right\}$ and $\left\{\mathcal{E}_i^{O^{s,o'}}\right\}$ related respectively to the points $\mathcal{W}_{obj} = \left[w_{obj_1},\ldots,w_{obj_4}\right]$ and $\mathcal{W}_{obj}^{'} = \left[w_{obj_1}^{'},\ldots,w_{obj_4}^{'}\right]$.
   %These sequences encapsulate respectively the obstacles $Ob^{o}_1$ and $Ob^{o'}_1$ (e.g. as in Figure \ref{fig:obstacle} where $\left\{\mathcal{E}_i^{s,O^{o'}}\right\}$ is depicted) ; 
		%\item the switching sequence $\left\{\mathcal{E}_i^{s,SW^{o'}}\right\}$ related to the way-points $\mathcal{W}_{sw} = \left[w_{sw_1},\ldots,w_{sw_2}\right]$ that is devoted to ensure a safe transition from the current state to the new sequence $\left\{\mathcal{E}_i^{s,o'}\right\}$ when an obstacle scenario change occurs (see Figure \ref{fig:switching}).
\end{itemize}

\subsection{Set-Theoretic RHC for Obstacle Avoidance and Motion Planning}
%
%With reference to Figure \ref{fig:waypoints}, given:
%\begin{itemize}
	%\item the set of available way-points ($\mathcal{W} = \left[w_1,\ldots,w_6\right]$) characterizing a feasible path (green trajectory);
    %\item the set of available way-points ($\mathcal{W}' = \left[w_7,\ldots,w_6\right]$) characterizing an alternative feasible path (red trajectory);
	%\item the set of available points $\mathcal{W}_{obj} = \left[w_{obj_1},\ldots,w_{obj_4}\right]$ and $\mathcal{W}_{obj}^{'} = \left[w_{obj_1}^{'},\ldots,w_{obj_4}^{'}\right]$ describing the obstacles to be avoided;
	%\item the set of available way-points ($\mathcal{W}_{sw} = \left[w_{sw_1},\ldots,w_{sw_2}\right]$) referred to a switching sequence allowing the vehicle to move between feasible paths;
%\end{itemize}
The aim is to design the final full path (e.g. the red arrows in Figure \ref{fig:scenario_change2}) that enable the vehicle to reach a destination goal. To achieve this, inspired by \cite{OAMP}, the accounted approach relies on a strategy that exploits% precomputed `safe zones', represented as sets of controllable states, where the vehicle can move without risk. 
%i = scenario di ostacolo
%More specifically, this approach consists of 
three key steps: 
\begin{enumerate}
	\item computing safe controllable zones (\textbf{Algorithm 1}),
	\item designing a viable path \textbf{(\textbf{Algorithm 2})},
	\item adapting the route dynamically in response to obstacles.
\end{enumerate}

The last step captures real-world conditions that are constantly changing and require real-time navigation adjustments. It enables real-time monitoring of the vehicle's position, verifying whether it remains on a valid and safe path. If a new obstacle appears or the current trajectory becomes unfeasible, the vehicle must dynamically shift to an alternative route, smoothly adapting to unexpected situations without compromising safety.
%This procedure is summarized in \textbf{Algorithm 3}.

To formally execute this strategy, an online Receding Horizon Control (RHC) procedure is required to select the optimal control input at each time step. Before outlining the algorithm, it is essential to define its key operational components. At any time $t$, the controller must first identify the most appropriate robust invariant safe region containing the current vehicle state $x(t)$. This is achieved by searching for the pair of indices $(s(t), i(t))$ - where $s$ denotes the specific path segment family and $i$ indicates the specific ellipsoid within that family - that satisfies the set-membership condition. This search is performed in lexicographic order to prioritize forward progress along the planned route. Once the active ellipsoid is identified, the system evaluates whether the vehicle has reached the local equilibrium point ($i(t) = 0$). If not, a control action $u(t)$ must be computed to drive the state into the next, smaller inner ellipsoid $\mathcal{E}_{i(t)-1}^{s(t)}$. To ensure rapid progression, the controller minimizes a specific running cost formulated as a min-max optimization problem over the system's polytopic vertices $j=1,\ldots,k$. The cost function utilizes the inverse of the shaping matrix, $(P_{i(t)-1}^{s(t)})^{-1}$, which characterizes the target ellipsoid $\{x \in \mathbb{R}^n : (x-x_{eq}^{s(t)})^T (P_{i(t)-1}^{s(t)})^{-1} (x-x_{eq}^{s(t)}) \le 1\}$. Minimizing this norm effectively pushes the predicted state as deeply as possible into the target region, approximating a one-step minimum-time control action \cite{angeli1}. The complete online procedure is summarized in \textbf{Algorithm 3}.

\noindent \rule{\columnwidth}{0.05cm}
\noindent \textbf{Algorithm 3 - OAMP Algorithm}\\
\noindent \rule{\columnwidth}{0.05cm}
\begin{enumerate}
	\item[(1)] $t=0$;
	\item[(2)] Solve in lexicographic order with respect to $s$ and $i$
	%\vspace{-0.3cm}
	\normalsize
\[
(s(t),i(t)) = \arg  \! \! \! \! \! \!    \min_{\substack {s \in \left\{0,\dots,N_s\right\} \\i \in  \left\{0,\dots,N_\mathcal{E}^s\right\}}} \! \! \! \left\{ \mathcal{A}_1 \right\}
\]
with: $$\mathcal{A}_1 = (s,i) : (x(t) - x^s_{eq})^T (P^{s}_i)^{-1} (x(t) - x^s_{eq}) \leq 1$$
\normalsize
	\item[(3)] \textbf{If} $i(t)=0$ \textbf{then} $u(t) = K^{s(t)} (x(t) - x^{s(t)}_{eq})$;
	\item[(4)] \textbf{Else}
	\begin{description}
	\item $ \begin{array}{cl}
	u(t) = \mbox{argmin}_u \mbox{max}_j& \! \! \! \! \! \!  \left\| \mathcal{B} \right\|^2_{(P^{s(t)}_{i(t)-1})^{-1}}
			\end{array}$
	\item \hspace{1cm} subject to	
	\item   \vspace{0cm}$$ \begin{array}{c}
	\mathcal{B} \in  \mbox{In} \left[ \mathcal{E}^{s(t)}_{i(t)-1} \sim G_{d_i} \mathcal{D} \right],\\ 
	u \in \mathcal{U};
	\end{array}$$
	\item $j=1,\ldots,k$
	\item with:
	$$
	\mathcal{B} = \Phi_j (x - x^{s(t)}_{eq}) + G_j(u - u^{s(t)}_{eq})
	$$
	\end{description}
	\normalsize
	\item [(5)] Apply $u(t)$; $t=t+1$; go to step (2);
\end{enumerate}
\noindent \rule{\columnwidth}{0.05cm}
\\
%Note that the running cost $$\mbox{max}_j\left\|\Phi_j (x-x^{s(t)}_{eq})+G_j (u-u^{s(t)}_{eq})\right\|^2_{(P^{s(t)}_{i(t)-1})^{-1}}$$ approximates  the one-step minimum-time control action \cite{angeli1} with $P^{s(t)}_{i(t)-1}$ denoting the shaping matrix of the ellipsoidal region $\left\{x \in \mathbb{R}^n:(x-x^{s(t)}_{eq})^T(P^{s(t)}_{i(t)-1})^{-1}(x-x^{s(t)}_{eq})\leq 1\right\}$.\\

This online procedure ensures that the vehicle safely navigates obstacles while remaining within feasible and controllable regions, leveraging the precomputed sequences to guide decision-making in real-time. For further analytical details refers to \cite{OAMP}.

%\begin{remark}
%It is important to highlight that, \textbf{Algorithm 1} and \textbf{2} are used to determine offline the sequences $\left\{\mathcal{E}^{s,o}\right\}$, $\left\{\mathcal{E}_s^{o'}\right\}$, $\left\{\mathcal{E}_s^{SW^{o'}}\right\}$, $\left\{\mathcal{E}_s^{O^o}\right\}$, $\left\{\mathcal{E}_s^{O^{o'}}\right\}$, the equilibrium points $\left\{x_{eq}^o\right\}$, $\left\{x_{eq}^{SW^{o'}}\right\}$, $\left\{x_{eq}^{O^o}\right\}$, $\left\{x_{eq}^{O^{o'}}\right\}$, the feedback gains $\left\{K^o\right\}$, $\left\{K^{o'}\right\}$, $\left\{K_{s}^{SW^{o'}}\right\}$, $\left\{K_{s}^{O^o}\right\}$, $\left\{K_{s}^{O^{o'}}\right\}$ respectively related to the set of points $\mathcal{W}$, $\mathcal{W}'$, $\mathcal{W}_{sw}$, $\mathcal{W}_{obj}$, $\mathcal{W}_{obj}'$.
%\end{remark}

%\begin{remark}
\textit{Remark 2 -  
    It is important to note that, the procedure described above, refereed to the particular case reported in Figure \ref{fig:waypoints}, can be easily extended to the any general driving scenario.}
%\end{remark}

%\begin{equation}
%\begin{array}{ll}
%	\mathcal{E}_{wp}^\ast & = \left( \underset{\begin{array}{c}
%					\forall i\\
%					s
%				\end{array}}{\bigcup}\mathcal{E}_s^i\right) \cup \left(\underset{\begin{array}{c}
%					\forall i \\
%					s
%				\end{array}}\bigcup\mathcal{E}_s^{O^i} \right) \\
%				&\cup \left(\underset{\begin{array}{c}
%					\forall i,s \\
%					s
%				\end{array}}\bigcup\mathcal{E}_s^{SW^i} \right)
%			\end{array}			
%\label{eq:single_set}
%\end{equation}

%%%%%%

%%%%%%
\section{Vehicle Model}
\label{s3}
This section details the vehicle dynamics modeling. 
To bridge the gap between rigorous control design and practical real-time implementation, a dual-model architecture is used. First, a high-fidelity 6-DOF nonlinear model \cite{r21} is introduced (Figure \ref{fig:modello4ruote}). This model acts as a "virtual plant" to simulate the real-world vehicle physics during validation, testing the controller against complex unmodeled dynamics.
Second, a simplified bicycle model describing planar motion is utilized (Figure \ref{fig:bicycle_model}). Moreover, this simplified model is embedded into an uncertain polytopic system and used strictly to construct the sequences of robustly control-invariant ellipsoidal sets and local feedback gains during the design phase. By separating the high-fidelity simulation model from the simplified control-oriented model, the proposed approach ensures mathematical robustness while maintaining the low computational footprint required for real-time execution For geometric parameters, forces, and reference frames, see Figures \ref{fig:modello4ruote} and \ref{fig:bicycle_model}.

%A 6-DOF nonlinear model is employed for validation, while a simplified bicycle model describing planar motion supports the model-based control design. To handle overtaking maneuvers, longitudinal dynamics are characterized via relative velocities between vehicles. Finally, to account for model uncertainties in the control strategy, an uncertain polytopic approximation of the nonlinear system is derived using an embedding approach. For geometric parameters, forces, and reference frames, see Figures \ref{fig:modello4ruote} and \ref{fig:bicycle_model}.
%
\subsection{6-DOF Non-Linear Vehicle Model}
%
%A 6-DOF nonlinear model \cite{r21}, accounting for longitudinal, lateral, yaw, and wheel dynamics, is adopted to derive a state-space representation suitable for control design.

%
\begin{figure}
	\centering
	\includegraphics[width=1\linewidth]{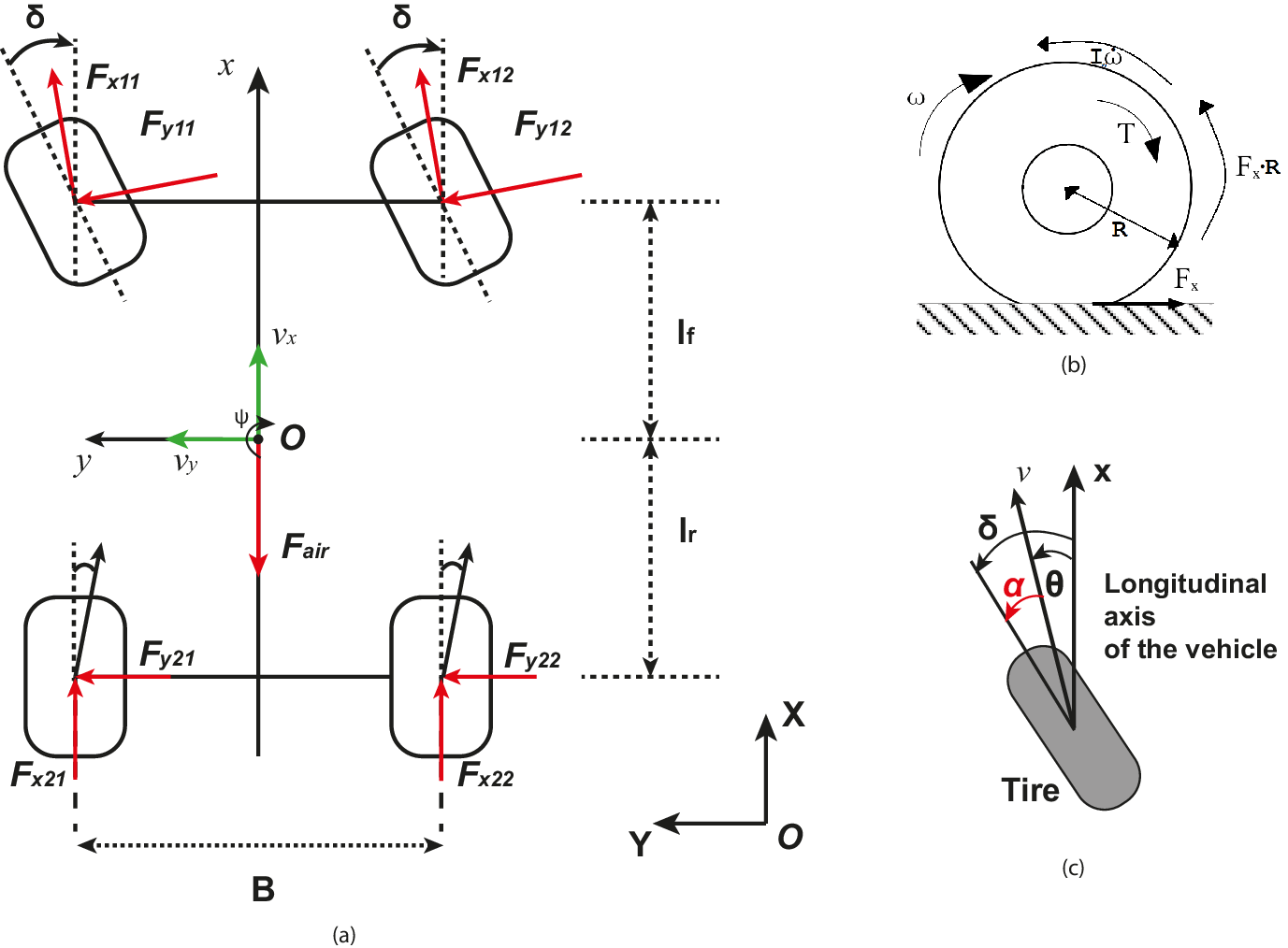}
	\caption{Vehicle Dynamic Model (a), Wheel Dynamic Model (b) and Slip Angle $\alpha$ (c).}
	\label{fig:modello4ruote}
\end{figure}
Figure \ref{fig:modello4ruote}(a) illustrates the body ($x0y$) and world ($X0Y$) reference frames. According to Newton's second law, the vehicle's longitudinal, lateral, and yaw dynamics take the following form:
\small
\begin{equation}
\left\{\begin{array}{lcl}
m \ddot{\mathbf{x}} & = & F_{x21}+F_{x22}+(F_{x11}+F_{x12})\cos(\delta) +\\
 & &- (F_{y11}+F_{y12})\sin(\delta) + m \dot{\psi} \mathbf{\dot{y}} - F_{air} \\[6pt]
m \ddot{\mathbf{y}} & = & F_{y21}+F_{y22}+(F_{x11}+F_{x12})\sin(\delta) +\\
 & &+ (F_{y11}+F_{y12})\cos(\delta) - m \dot{\psi} \mathbf{\dot{x}} \\[6pt]
I_z \ddot{\psi} & = & [(F_{x11}+F_{x12})\sin(\delta) +(F_{y11}+F_{y12})\cos(\delta)]l_f +\\
& &+[(F_{x12}-F_{x11})\cos(\delta)+(F_{y11}-F_{y12})\sin(\delta)]\frac{B}{2} + \\
& &-(F_{y21}+F_{y22})l_r+(F_{x22}-F_{x21})\frac{B}{2}
\end{array}\right.
\label{eq1}
\end{equation}
\normalsize
Note that, the index  $i$ indicates the axes ($i=1$ for the front axis and $i=2$ for the rear axis) whilst $j$ indicates the wheel ($j=1$ for the left wheels and $j=2$ for the right wheels);
The body coordinates $\dot{\mathbf{x}}$, $\dot{\mathbf{y}}$ and $\psi$ are then transformed into the world coordinates $\dot{X}$ and $\dot{Y}$ as follows
\begin{equation}
\left\{\begin{array}{lcl}
\dot{X} & = & \dot{\mathbf{x}}\cos(\psi)-\dot{\mathbf{y}}\sin(\psi) \\[6pt]
\dot{Y} & = & \dot{\mathbf{x}}\sin(\psi)+\dot{\mathbf{y}}\cos(\psi)
\end{array}\right.
\end{equation}
Moreover, with reference to Figure \ref{fig:modello4ruote}(b), it is possible to obtain the dynamic models of the front and rear wheel rotations \footnote{We consider a front-wheel drive vehicle}  as \cite{r21}
\begin{equation}
\left\{\begin{array}{lcl}
I_w \dot{\omega}_{1j} & = & T_{1j}-F_{x1j}R \quad j=1,2 \\ [6pt]
I_w \dot{\omega}_{2j} & = & -F_{x2j}R \quad j=1,2
\end{array}\right.
\label{rolling1}
\end{equation}
where also in this case $j$ indicates the wheel.
The longitudinal and lateral forces on the tire can be computed using Pacejka's Magic Formula \cite{r21}. In this context, it is essential to introduce the concepts of slip angle and slip ratio. Referring to Figure \ref{fig:modello4ruote}(c), the slip angle for each wheel is defined as follows:
\small
\begin{equation}\label{slipangle}
\left\{\begin{array}{l}
\alpha_{11}=\delta - \arctan\left(\frac{\mathbf{\dot{y}} + l_f \dot{\psi}}{\mathbf{\dot{x}} - \frac{B \dot{\psi}}{2}}\right), \ \ \alpha_{12}=\delta - \arctan\left(\frac{\mathbf{\dot{y}} + l_f \dot{\psi}}{\mathbf{\dot{x}} + \frac{B \dot{\psi}}{2}}\right)\\[8pt]
\alpha_{21}=\arctan\left(\frac{l_r \dot{\psi} - \mathbf{\dot{y}}}{\mathbf{\dot{x}} - \frac{B\dot{\psi}}{2}}\right), \ \ \alpha_{22}=\arctan\left(\frac{l_r \dot{\psi} - \mathbf{\dot{y}}}{\mathbf{\dot{x}} + \frac{B\dot{\psi}}{2}}\right)
\end{array}\right.
\end{equation}
\normalsize
On the other side, the slip ratio is defined as
\begin{equation}\label{slipratio}
\sigma_{xij} = \begin{cases}
\frac{R \omega_{ij} - \mathbf{\dot{x}}_{ij}}{R \omega_{ij}}, \, R\omega > \mathbf{\dot{x}} \, (Acceleration)\\
\frac{R \omega_{ij} - \mathbf{\dot{x}}_{ij}}{\mathbf{\dot{x}}_{ij}}, \, R\omega < \mathbf{\dot{x}} \, (Braking)
\end{cases}, i,j=1,2 
\end{equation}
where the longitudinal velocity of each wheel ($\mathbf{\dot{x}}_{ij}$) is defined as
\begin{equation}
\left\{
\begin{array}{l}
	\mathbf{\dot{x}}_{11} = \left(\mathbf{\dot{x}} - \frac{B \dot{\psi}}{2}\right)\cos \delta + \left(\mathbf{\dot{y}} + l_f \dot{\psi}\right)\sin \delta\\ [6pt]
	\mathbf{\dot{x}}_{12} = \left(\mathbf{\dot{x}} + \frac{B \dot{\psi}}{2}\right)\cos \delta + \left(\mathbf{\dot{y}} + l_f \dot{\psi}\right)\sin \delta \\[6pt]
	\mathbf{\dot{x}}_{21} = \mathbf{\dot{x}} - \frac{B \dot{\psi}}{2} \\[6pt]
	\mathbf{\dot{x}}_{22} = \mathbf{\dot{x}} + \frac{B \dot{\psi}}{2}
\end{array}
\right.
\label{eq:speedwheel1}
\end{equation}
The mathematical expression of the tire forces exchanged with the road  is given by the sinusoidal version of the Magic Formula \cite{c11}
\begin{equation}
F(k)=D \sin (C \arctan(S k - E (S k - \arctan(S k))))+ S_{V}
\label{magic}
\end{equation}
with
\begin{equation}
k = K + S_{H}
\end{equation}
In eq. (\ref{magic}), depending on the type of motion considered (longitudinal or lateral), $F$ represents either the longitudinal force $F_x$ or the lateral force $F_y$, while $K$ represents the Slip Ratio $\sigma_x$ or the Slip Angle $\alpha$.
Note that the parameters of the Magic Formula can be found in \cite{r21}.
%Note that Table \ref{symbol} summarizes the main symbols and units of the 6-DOF model, whereas the parameters of the Magic Formula can be found in \cite{r21}.

\subsection{Vehicle bicycle model}
\label{s3_1}
\begin{figure}[htbp]
	\centering
		\includegraphics[width=0.4\textwidth]{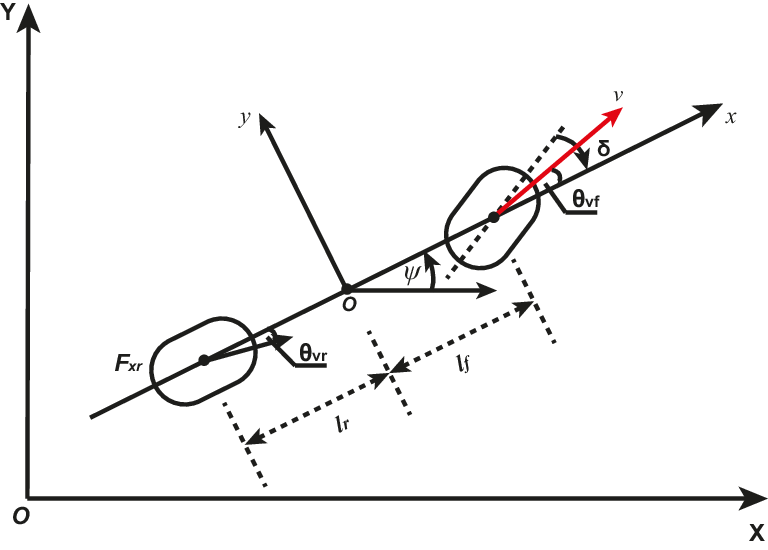}
	\caption{Vehicle lateral bicycle model. ($x,y$) body coordinates, (\textbf{X},\textbf{Y}) world coordinates.}
	\label{fig:bicycle_model}
\end{figure}

According to Figure \ref{fig:bicycle_model}, it is possible to derive a three degrees of freedom vehicle bicycle model, where the degrees of freedom are represented by the vehicle longitudinal ($\mathbf{x}$) and lateral ($\mathbf{y}$) positions and the vehicle yaw angle ($\psi$) \cite{r29}. In particular, the longitudinal, lateral and yaw dynamics can be obtained starting from Newton's second law for motion along the $\mathbf{x}$-axis, $\mathbf{y}$-axis and by balancing the moments around the vertical $\mathbf{z}$-axis:
	\begin{eqnarray}
	%\left\{\begin{array}{lcl}
			m\mathbf{\ddot{x}} & = & F_{xf}+F_{xr} \label{fb1}\\
				m\mathbf{\ddot{y}} & = & F_{yf}+F_{yr} \label{fb2}\\
				I_z\ddot{\psi}& = & l_fF_{yf}-l_rF_{yr} \label{fb3}
	%\end{array}\right.
		%\label{fb}
	\end{eqnarray}
where, $m$ and $I_z$ denote the vehicle mass and yaw inertia, $l_f, l_r$ the distances from the center of gravity (CoG) to the front/rear axles, while $F_{i,j}$ (with $i \in \{x,y\}, j \in \{f,r\}$) represents the longitudinal/lateral tire forces. Finally, $\dot{\psi}$ is the yaw rate, and $\delta$ is the steering angle.
It is important to note that, in eq. (\ref{fb2}) two terms contribute to the lateral acceleration ($\mathbf{a}_y$):
\begin{enumerate}
	\item $\ddot{\mathbf{y}}$, the acceleration due to the motion along the $\mathbf{y}$-axis;
	\item $\mathbf{\dot{x}}\dot{\psi}$, the centripetal acceleration.
\end{enumerate}
Moreover, the lateral tire forces are proportional to the slip angle, which is defined as the angle between the tire's orientation and the direction of the wheel's velocity vector (see Figure \ref{fig:modello4ruote}(c)).
%
%\begin{figure}[htbp]
%	\centering
%		\includegraphics[width=0.2\textwidth]{Figure/tire_model.eps}
%	\caption{Tire slip angle $\alpha$.}
%	\label{fig:tire_model}
%\end{figure}
%
Furthermore, by introducing:
\begin{itemize}
	\item the steering angle $\delta$;
	\item the longitudinal velocity $\mathbf{\dot{x}}$;
	\item the Cornering Stiffness parameters $C_{\alpha f}$ and $C_{\alpha r}$;
	\item the front and rear tire velocity angle $\theta_{vf}$ and $\theta_{vr}$;
\end{itemize}
and, by assuming a small-angle approximation \cite{c2} that allows one to define
\begin{equation}
\theta_{vf} \approx \frac{\dot{\mathbf{y}}-l_f\dot{\psi}}{\mathbf{\dot{x}}}, \, \, \, \, \, \, \theta_{vr} \approx \frac{\dot{\mathbf{y}}-l_r\dot{\psi}}{\mathbf{\dot{x}}}
\label{small_angle}
\end{equation}
Then, eqs. (\ref{fb2})-(\ref{fb3}) can be written as
\small
\begin{equation}
\left\{
	\begin{array}{rcl}
		m(\ddot{\mathbf{y}}+\dot{\psi}\mathbf{\dot{x}}) &=&\underbrace{2C_{\alpha f}(\delta-\theta_{vf})}_{F_{yf}}+\underbrace{2C_{\alpha f}(-\theta_{vr})}_{F_{yr}} \\[8pt]
		I_z\ddot{\psi} &=& l_f\underbrace{2C_{\alpha f}(\delta-\theta_{vf})}_{F_{yf}}-l_r\underbrace{2C_{\alpha f}(-\theta_{vr})}_{F_{yr}}
	\end{array}
\right.
\label{eq_4}
\end{equation}
that leads to 
\small
	\begin{equation}
		\left\lbrace
		\begin{array}{lcl}
			\ddot{\mathbf{y}}&\! \! \! = \! \! \!& \left(-\mathbf{\dot{x}}-\dfrac{2C_{\alpha f}l_f-2C_{\alpha r}l_r}{m\mathbf{\dot{x}}}\right) \dot{\psi} +\left(-\dfrac{2C_{\alpha f}+2C_{\alpha r}}{m\mathbf{v}_x}\right)\dot{y} +\\
			&&\left(\dfrac{2C_{\alpha f}}{m}\right)\delta\\[10pt]
			\ddot{\psi}&\! \! \!=\! \! \!&\left(-\dfrac{2l_f^2C_{\alpha f}-2l_r^2C_{\alpha r}}{I_z\mathbf{\dot{x}}}\right) \dot{\psi}+\left(-\dfrac{2l_fC_{\alpha f}-2l_rC_{\alpha r}}{I_z\mathbf{\dot{x}}}\right) \dot{\mathbf{y}} + \\
			&& \left( \dfrac{2l_fC_{\alpha f}}{I_z}\right) \delta \\
		\end{array}
		\right.
		\label{lateral}
	\end{equation}
\normalsize	
Note also that the lateral position $\mathbf{y}$ expressed in the body coordinates can be transformed in the world coordinates as follows
\begin{equation}
		\dot{Y}=\dot{\mathbf{y}}\cos(\psi)+\mathbf{\dot{x}}\sin(\psi),
\end{equation}
that, due to small angle approximation, simplifies to 
\begin{equation}
		\dot{Y}\approx\dot{\mathbf{y}}+\mathbf{\dot{x}}\psi.
		\label{worldY}
	\end{equation}
In addition, to rigorously analyze lateral motion in autonomous driving scenarios, such as overtaking, it is necessary to introduce the lateral relative distance concept. Specifically, the variables $\mathbf{y}^{rel}$ and $\dot{\mathbf{y}}^{rel}$ represent, respectively, the lateral distance and speed difference between the current vehicle (ego car) and the preceding vehicle (lead car):
	\begin{equation}
		\mathbf{y}^{rel}=\mathbf{Y}^{ego}-\mathbf{Y}^{lead},
		\label{yrel}
\end{equation}
\begin{equation}
\begin{array}{lcl}
		\dot{\mathbf{y}}^{rel} &= & \dot{\mathbf{Y}}^{ego}-\dot{\mathbf{Y}}^{lead}= \dot{\mathbf{y}}+\mathbf{\dot{x}}\psi -\dot{\mathbf{Y}}^{lead}.
		\label{vyrel}
		\end{array}
\end{equation}
 where: 
\begin{itemize} 
\item $\mathbf{Y}^{ego}$ and $\dot{\mathbf{Y}}^{ego}$ denote the lateral position and speed of the ego car; 
\item $\mathbf{Y}^{lead}$ and $\dot{\mathbf{Y}}^{lead}$ denote the lateral position and speed of the lead car. 
\end{itemize}
%Furthermore, the lateral speeds $\dot{\mathbf{y}}^{ego}$ and $\dot{\mathbf{y}}^{lead}$ can be expressed as: 
%\begin{equation}
%		\dot{\mathbf{y}}^{ego}=\bar{\mathbf{v}}_y+\Delta \mathbf{v}_y^{ego}; \qquad \dot{\mathbf{y}}^{lead}=\bar{\mathbf{v}}_y+\Delta \mathbf{v}_y^{lead};
%\label{speedsy}
%\end{equation}
% where $\bar{\mathbf{v}}_y$ is the nominal lateral velocity, and $\Delta \mathbf{v}_y^{ego}=\mathbf{\dot{y}}$ and $\Delta \mathbf{v}_y^{lead}$ represent deviations from the nominal values.
%Substituting expressions from (\ref{speedsy}) into (\ref{vyrel}) yields: 
%\begin{equation}
%\begin{array}{lcl}
%	\dot{\mathbf{y}}^{rel} \! \! \! &=& \! \! \dot{\mathbf{y}}^{ego}-\dot{\mathbf{y}}^{lead}=\left( \bar{\mathbf{v}}_y+\Delta \mathbf{v}_y^{ego}\right) -\left(\bar{\mathbf{v}}_y+\Delta \mathbf{v}_y^{lead}\right)= \\
%	\! \! \! &=& \! \!\Delta \mathbf{v}_y^{ego}-\Delta \mathbf{v}_y^{lead}. 
%\end{array} 
%\label{new_speedy} 
%\end{equation}
 %This formulation highlights that the relative lateral velocity depends solely on the variations in the vehicles' lateral speeds.
%
\subsection{Longitudinal vehicle dynamics}
\label{s3_2}
%
%
%\begin{figure}[htbp]
%	\centering
%		\includegraphics[width=0.5\textwidth]{Figure/two_vehicle.eps}
%	\caption{Two vehicles car-following model.}
%	\label{fig:two_vehicle}
%\end{figure}
%

%
The characterization of the vehicle's longitudinal motion, can be done by accounting the following simplified equation
\begin{equation}
		\ddot{\mathbf{x}}= \dot{\Delta \mathbf{v}}_x = \dot{\mathbf{y}}\dot{\psi}+\ddot{\mathbf{x}}^{des},\\
		\label{longitudinal}
\end{equation}
which implies that the vehicle longitudinal acceleration $\ddot{\mathbf{x}}$ can be obtained as the sum of two contributions:
\begin{itemize}
	\item $\ddot{\mathbf{x}}^{des}$, the desired vehicle acceleration  (e.g. due to a driving or braking force imposed by a control system)
	\item $\dot{\mathbf{y}}\dot{\psi}$, the coupling term representing the effect of lateral dynamics on longitudinal dynamics.
\end{itemize}
Note also that the lateral position $\mathbf{x}$ expressed in the body coordinates can be transformed in the world coordinates as follows
\begin{equation}
		\dot{X}=\dot{\mathbf{x}}\cos(\psi)-\mathbf{v}_y\sin(\psi),
\end{equation}
that, due to small angle approximation, simplifies to 
\begin{equation}
		\dot{X}\approx\dot{\mathbf{x}}-\mathbf{\dot{y}}\psi.
		\label{worldX}
	\end{equation}
Moreover, to better analyze the vehicle longitudinal motion in autonomous vehicle's context (e.g. an overtaking maneuver), it is necessary to introduce the longitudinal relative distance $\mathbf{x}_{rel}$ and velocity $\dot{\mathbf{x}}_{rel}$, that are the distance and the speed difference between the current vehicle (named \textit{ego car}) and a preceding vehicle (named \textit{lead car}):
\begin{equation}
		\mathbf{x}^{rel}=\mathbf{X}^{ego}-\mathbf{X}^{lead},
		\label{xrel}
\end{equation}
\begin{equation}
		\dot{\mathbf{x}}^{rel}= \dot{\mathbf{X}}^{ego}-\dot{\mathbf{X}}^{lead}.
		\label{vrel}
\end{equation}
where:
\begin{itemize}
	\item $\mathbf{X}^{ego}$ and $\dot{\mathbf{X}}^{ego}$ represent the longitudinal position and speed of the ego car;
	\item $\mathbf{X}^{lead}$ and $\dot{\mathbf{X}}^{lead}$ are the longitudinal position and speed of the lead car.
\end{itemize}
In addition, the longitudinal speeds $\dot{\mathbf{x}}^{ego}$ and $\dot{\mathbf{x}}^{lead}$ can be expressed as 
\begin{equation}
		\dot{\mathbf{x}}^{ego}=\bar{\mathbf{v}}_x+\Delta \mathbf{v}_x^{ego}; \qquad \dot{\mathbf{x}}_x^{lead}=\bar{\mathbf{v}}_x+\Delta \mathbf{v}_x^{lead};
\label{speeds}
\end{equation}
where $\bar{\mathbf{v}}_x$ is the nominal velocity and $\Delta \mathbf{v}_x^{ego}$ and $\Delta \mathbf{v}_x^{lead}$ represent a variation on vehicles cruise velocities.
By combining the two equations in (\ref{speeds}), the relative velocity can be rewritten as
\begin{equation}
\begin{array}{lcl}
	\dot{\mathbf{x}}^{rel} &=& \dot{\mathbf{x}}^{ego}-\dot{\mathbf{x}}^{lead}=\left( \bar{\mathbf{v}}_x+\Delta \mathbf{v}_x^{ego}\right) -\left(\bar{\mathbf{v}}_x+\Delta \mathbf{v}_x^{lead}\right) \\
	&=&\Delta \mathbf{v}_x^{ego}-\Delta \mathbf{v}_x^{lead}. 
\end{array} 
\label{new_speed} 
\end{equation}
This formulation highlights that the relative velocity depends only on vehicles' velocity variations.
\subsection{Non-linear state-space representation}
\label{s3_3}
By considering eqs. (\ref{lateral}), (\ref{vyrel}),(\ref{longitudinal}) and (\ref{new_speed}) and the systems vectors
\begin{itemize}
	\item $x = \left[\Delta \mathbf{v}_x^{ego}, \dot{y}, \psi, \dot{\psi}, \mathbf{y}^{rel},  \mathbf{x}^{rel} \right]^T \in \mathbb{R}^{6\times1}$, the plant state;
	\item $u = \left[\delta_f, \ddot{\mathbf{x}}^{des}\right]^T \in \mathbb{R}^{2\times1}$, the manipulable inputs;
	\item $d = \left[\dot{Y}^{lead}, \Delta \mathbf{v}_x^{lead}\right] \in \mathbb{R}^{2\times 1}$, the disturbance;
	\item $y = \left[\dot{\mathbf{x}}, \dot{\mathbf{y}}, \psi, \dot{\psi}, \mathbf{y}^{rel},  \mathbf{x}^{rel} \right]^T \in \mathbb{R}^{6\times1}$, the measured output;
\end{itemize}
the following non-linear state-space representation of the longitudinal and lateral vehicle dynamics can be achieved

	\begin{equation}
		\left\lbrace
		\begin{array}{lcl}
			\dot{x}_1&=&x_2x_4+u_2\\[8pt]
			\dot{x}_2&=&\left(-\dfrac{2C_{\alpha f}+2C_{\alpha r}}{M\left(\bar{\mathbf{v}}+x_1\right) }\right) x_2 + \left(\dfrac{2C_{af}}{M}\right)u_1+ \\
			&&\left(-\left(\bar{v}+x_1\right) -\dfrac{2C_{\alpha f}l_f-2C_{\alpha r}l_r}{M\left( \bar{\mathbf{v}}+x_1\right) }\right) x_4\\
			\dot{x}_3&=&x_4\\[8pt]
			\dot{x}_4&=&\left(-\dfrac{2l_fC_{\alpha f}-2l_rC_{\alpha r}}{I_z\left(\bar{\mathbf{v}}+x_1\right) }\right) x_2 + \left( \dfrac{2l_fC_{\alpha f}}{I_z}\right)u_1+ \\
			&&\left(-\dfrac{2l_f^2C_{\alpha f}-2l_r^2C_{\alpha r}}{I_z\left(\bar{\mathbf{v}}+x_1\right) }\right) x_4\\
			\dot{x}_5&=&x_2+\left(\bar{\mathbf{v}}+x_1\right)x_3-d_1\\[8pt]
			%\dot{x_6}&=&\left(\bar{\mathbf{v}}+x_1\right)-x_2x_3\\
			\dot{x}_6&=& x_1-d_2%\\
			%y_1 & = & x_1 \\
			%y_2 & = & x_2 \\
			%y_3 & = & x_3 \\
			%y_4 & = & x_4 \\
			%y_5 & = & x_5 \\
			%y_6 & = & x_6
		\end{array}
		\right.
		\label{STM}
	\end{equation}
\normalsize
\textit{Remark 3 -  
Note that, the exogenous disturbance vector $d$ explicitly models the real-world, dynamic, and unpredictable actions of the surrounding traffic (i.e., the dynamic obstacles) with the possibility to include sensor noise as well. Specifically, it quantifies the variations in the lead vehicle's longitudinal speed and lateral position.
}
%\vspace{-2cm}
%
\subsection{Embedded model}
\label{s4}

\begin{figure*}[t]
\footnotesize
	\begin{equation}
		\Phi=\left[ 
		\begin{array}{cccccc}
			0 & \gamma_2 & 0 & 0 & 0 & 0\\
			-\gamma_2 & -\gamma_3\dfrac{2C_{\alpha f}+2C_{\alpha r}}{M} & 0 & 
			-\left(\bar{\mathbf{v}} +\dfrac{2C_{\alpha f}l_f-2C_{\alpha r}l_r}{M}\gamma_3\right) & 0 & 0\\
			0 & 0 & 0 & 1 & 0 & 0\\
			0 & -\dfrac{2l_fC_{\alpha f}-2l_rC_{\alpha r}}{I_z}\gamma_3 & 0 & 
			-\dfrac{2l_f^2C_{\alpha f}-2l_r^2C_{\alpha r}}{I_z}\gamma_3 & 0 & 0\\
			\gamma_1 & 1 & \bar{\mathbf{v}} & 0 & 0 & 0\\
			1 & 0 & 0 & 0 & 0 & 0 \\
		\end{array}
		\right];
\quad
		G\left[ 
		\begin{array}{cc}
			0 							& 1				\\
			\dfrac{2C_{\alpha f}}{M}		 	& 0 			\\
			0 							& 0 			\\
			\dfrac{2l_fC_{af}}{I_z} 		& 0 			\\
			0 							& 0				\\
			0							& 0	
		\end{array}
		\right]; \quad
		\begin{array}{c}
		G_d\left[ 
		\begin{array}{cc}
			0 & 0 \\
			0 & 0 \\
			0 & 0 \\
			0 & 0 \\
			-1 & 0 \\
			0 &-1
		\end{array}
		\right]; \\
		\\
		C = I_{6 \times 6}.
		\end{array}
		\label{ssmatrices}
	\end{equation}
\end{figure*}
\normalsize
The control objective is to execute an overtaking maneuver by regulating the steering angle ($u_1$) and longitudinal acceleration ($u_2$). To enable a linear control design, the non-linear dynamics (\ref{STM}) are mapped into a convex polytopic structure via a Quasi-LPV (Linear Parameter-Varying) embedding. This procedure reformulates the non-linearities into a set of bounded, state-dependent scheduling parameters, yielding a pseudo-linear state-space realization suitable for robust ST-FE-MPC synthesis.
%Within the previously described context, the control objective is to enable the ego vehicle to execute an overtaking maneuver while adjusting its longitudinal speed through the regulation of the steering angle ($u_1$) and vehicle acceleration ($u_2$). To proceed with the linear control design, the non-linear dynamics (\ref{STM}) must be mapped into a convex polytopic structure. 
%This is achieved through a Quasi-LPV (Linear Parameter-Varying) embedding technique. The core idea of this procedure is to exactly reformulate the non-linear system by "hiding" the state-dependent non-linearities inside a set of bounded, time-varying scheduling parameters, thus obtaining a pseudo-linear state-space realization.
By analyzing the non-linear vehicle dynamics (\ref{STM}), it is evident that the non-linearities stem primarily from two sources: the multiplication between state variables (e.g., the coupling terms $x_2 x_4$) and the rational fractions caused by the time-varying longitudinal velocity term $1/(\bar{v} + x_1)$. To achieve an exact pseudo-linearization of the state-space equations without introducing approximation errors, we systematically isolate these specific non-linear terms and define the following time-varying scheduling parameters:
\begin{equation}
	\gamma_1 = x_3, \quad \gamma_2 = x_4, \quad \gamma_3 = 1/(\bar{\mathbf{v}}+x_1),
\label{eq:gamma}
\end{equation}
By algebraically substituting the definitions \eqref{eq:gamma} back into the non-linear equations, the state multiplications and rational fractions are factored out from the state vector. For instance, the non-linear coupling term $x_2 x_4$ in the longitudinal dynamics is factored as $\gamma_2 x_2$. This systematic substitution allows us to exactly rewrite the original non-linear system as the following parameter-dependent state-space realization $\dot{x}(t) = \Phi(\gamma(t))x(t) + G u(t) + G_d d(t)$:
\small
	\begin{equation}
		\left\lbrace
		\begin{array}{lcl}
			\dot{x}_1 \! \! \! \! \! &=& \! \! \! \! \! \gamma_2x_2+u_2\\[8pt]
			\dot{x}_2 \! \! \! \! \! &=& \! \! \! \! \!-\gamma_2x_1+\gamma_3\left(-\dfrac{2C_{\alpha f}+2C_{\alpha r}}{M}\right) x_2 +
			 \left(\dfrac{2C_{\alpha f}}{M}\right)u_1 - \\ 
			&& \left(\bar{\mathbf{v}}+\dfrac{2C_{\alpha f}l_f-2C_{\alpha r}l_r}{M}\gamma_3\right) x_4 \\[8pt]
			\dot{x}_3 \! \! \! \! \! &=& \! \! \! \! \! x_4\\[8pt]
			\dot{x}_4 \! \! \! \! \! &=& \! \! \! \! \! \gamma_3\left(-\dfrac{2l_fC_{\alpha f}-2l_rC_{\alpha r}}{I_z}\right) x_2 +\left( \dfrac{2l_fC_{\alpha f}}{I_z}\right)u_1+ \\
			&& \gamma_3\left(-\dfrac{2l_f^2C_{\alpha f}-2l_r^2C_{\alpha r}}{I_z}\right)x_4\\[8pt]
			\dot{x}_5 \! \! \! \! \! &=& \! \! \! \! \! \gamma_1x_1+x_2+\bar{\mathbf{v}}x_3-d_1\\[8pt]
			%\dot{x}_6&=&\left(\bar{v}+x_1\right) -x_2\gamma_1\\
			\dot{x}_6 \! \! \! \! \! &=& \! \! \! \! \! x_1-d_2%\\
			%y_1 & = & x_1 \\
			%y_2 & = & x_2 \\
			%y_3 & = & x_3 \\
			%y_4 & = & x_4 \\
			%y_5 & = & x_5 \\
			%y_6 & = & x_6
		\end{array}
		\right.
		\label{embLPV}
	\end{equation}
	\normalsize
With this formulation, the state-dependent non-linearities are completely embedded within the matrix $\Phi(\gamma)$. Consequently, by assuming that the parameters $\gamma_1, \gamma_2$, and $\gamma_3$ are bounded within known minimum and maximum operational limits:
\begin{equation}
\gamma_{i,min} \le \gamma_i(t) \le \gamma_{i,max}, \quad i=1,2,3
\end{equation}
the infinite-dimensional Quasi-LPV system \eqref{embLPV}, characterized by the system matrices (\ref{ssmatrices}), can be bounded by a finite convex polytope defined by its $2^3 = 8$ vertices by introducing the following auxiliary normalized function
\begin{equation}
\begin{array}{c}
	\rho_1 = \dfrac{\gamma_1-\gamma_{1 \ min }}{\gamma_{1 \ max}-\gamma_{1 \ min }}, \quad \rho_2 = \dfrac{ \gamma_2-\gamma_{2 \ min}}{\gamma_{2 \ max}-\gamma_{2 \ min }},\\[8pt]
	\rho_3 = \dfrac{ \gamma_3-\gamma_{3 \ min } }{\gamma_{3 \ max}-\gamma_{3 \ min }},% \quad \rho_4 = \dfrac{ \gamma_4-\gamma_{4 \ min}}{\gamma_{4 \ max}-\gamma_{4 \ min }},
\end{array}
\label{eq:normalized}
\end{equation}
that satisfies $0\leq \rho_i\leq 1$, $i=1,\ldots,3$ and are instrumentals to define the matrix vertices of the following uncertain polytopic model: 
\begin{equation}
		\left\lbrace 
		\begin{array}{l}
			\dot{x}=\Phi(p)x+Gu+G_dd\\
			y=Cx
		\end{array}
		\right.;
		\label{polytopic}
	\end{equation}
with 
\begin{equation}
p=\left[ p_1, \ldots,	p_{8} \right]^T
\label{eq:}
\end{equation}
and

	\begin{equation*}
	\begin{array}{lcl}
	\left[ 
	\begin{array}{ccc}
		\Phi(p) & G & G_d \\
		C & 0 & 0
	\end{array}
	\right]& \! \! \! \! \! \!:= \! \! \! \! \! \!&
	\left\lbrace 
	\begin{array}{l}
		\overunderset{l}{i=1}{\sum}p_i
	\left[
	\begin{array}{ccc}
		\Phi_i & G & G \\
		C & 0 & 0 
	\end{array}
	\right], \\
0 \le p_i \le 1, \overunderset{l}{i=1}{\sum}p_i=1
	\end{array}	
	\right\rbrace
	\end{array}	
		\end{equation*}
		\begin{equation}
		\in
		Co
		\left\lbrace
		\left[
		\begin{array}{ccc}
			\Phi_1 & G & G_{d}\\
			C & 0 & 0
		\end{array}
		\right],
		\ldots ,
		\left[
		\begin{array}{ccc}
			\Phi_8 & G & G_{d}\\
			C & 0 & 0
		\end{array}
		\right]
		\right\rbrace 
	\end{equation}
	
	\normalsize

\textit{Remark 4:} The parameters $\gamma_{1,2,3}$ treat intrinsic state-dependent nonlinearities as bounded, time-varying uncertainties. This polytopic structure encapsulates the complex vehicle kinematics and dynamics during maneuvers, rather than representing uncertainties in physical parameters (e.g., mass, friction) or external environmental factors.
%%%%%%%%%%%%%%%%%%%%%%%%%%%%%%%%%%%%%%%%%%
%%%%%%%%%%%%%%%%%%%%%%%%%%%%%%%%%%%%%%%%%%
%%%%%%%%%%%%%%%%%%%%%%%%%%%%%%%%%%%%%%%%%%

%%%
\section{Simulations}
\label{s6}
This section validates the proposed control strategy for safe, collision-free overtaking under dynamic constraints. The framework was initially implemented in MATLAB/Simulink using the Automated Driving Toolbox (ADT) to integrate sensor fusion and path planning with a 6-DOF nonlinear vehicle model. Subsequently, high-fidelity co-simulations were conducted in the CARLA 3D environment to rigorously test the controller against realistic physics, unmodeled dynamics, and complex traffic scenarios.
\begin{figure}[h]
		\centering
		\includegraphics[width=0.8\linewidth]{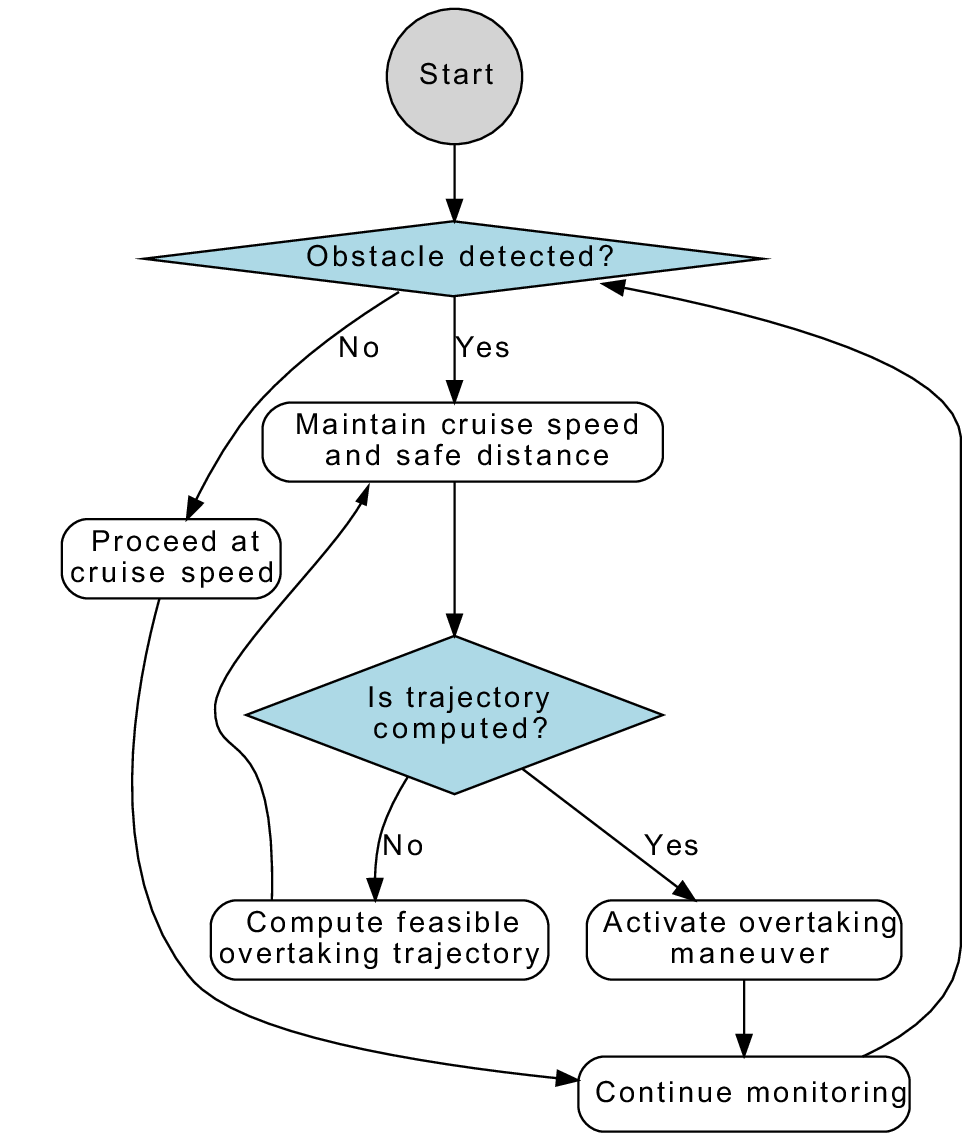}
				\caption{Control logic: flow chart.}
				\label{fig:flow_chart}
\end{figure}
\subsection{Simulation Framework}
\label{s6_0}
The 6-DOF model is discretized with a 0.1s sampling time. 
During the real-time closed-loop operation, the proposed Set-Theoretic MPC samples the current state of the 6-DOF virtual plant and evaluates it against the robust invariant ellipsoidal sequences (derived from the simplified model). Based on this set-membership evaluation, the algorithm computes the optimal steering angle $\delta_f$ and longitudinal acceleration $\dot{x}^{des}$, which are then fed-back into the 6-DOF model to dynamically update the vehicle's trajectory. 

An ADT-simulated radar enables the ego vehicle to maintain safe distances from preceding vehicles through adaptive speed regulation.  Moreover, the ego vehicle embedded polytopic model (\ref{polytopic}) was defined by the following parameter bounds:
\begin{equation}
\label{eq:boundsim}
\left\{ -\frac{\pi}{2} \leq \gamma_1 \leq \frac{\pi}{2}, \quad -2 \leq \gamma_2 \leq 2, \quad 0.03 \leq \gamma_3 \leq 0.1 \right.
\end{equation}
Furthermore, to guarantee robust obstacle avoidance, the sequence of invariant ellipsoidal sets was computed by assuming a bounded compact disturbance set $\mathcal{D}$. Specifically, the disturbances-representing the unpredictable lateral and longitudinal speed variations of the leading vehicles-were rigorously quantified and bounded as:
\begin{equation}
\begin{array}{l}
	\mathcal{D} := \left\{ d \in \mathbb{R}^2 : |\dot{Y}^{lead}| \le \text{0.5} \text{ [m/s]},
	 |\Delta v_x^{lead}| \le \text{1.5} \text{ [m/s]} \right\}
\end{array}
\end{equation}
These quantitative bounds ensure that the generated feasible trajectories remain strictly safe as long as the target vehicles maneuver within these velocity variation limits.
In what follows, two simulation scenario, referring to highway overtaking maneuver, are presented.
Since the objective is to ensure safety and smooth behavior during overtaking, to determine the state-feedback control law  (\ref{eq:control_law}), the following state and input constraints (\ref{eq:ux_constr}) are considered for the first simulation:

\begin{equation}
\label{eq:constraintsim}
\left\{\begin{aligned}
& -0.5 \leq u_1 \leq 0.5, \quad -2 \leq u_2 \leq 2, \quad -10 \leq x_1 \leq 10, \\
& -\frac{\pi}{2} \leq x_3 \leq \frac{\pi}{2}, \quad -2 \leq x_4 \leq 2, \quad -3 \leq x_5 \leq 3, \\
& \quad \quad \quad \quad \quad \quad \quad \quad 12 \leq |x_6| \leq 50.
\end{aligned}
\right.
\end{equation}
\begin{figure}[h!]
		\centering
		\includegraphics[angle=-90, width=0.8\linewidth]{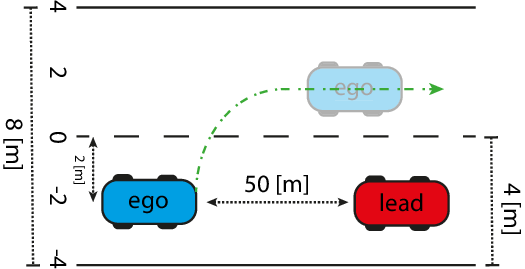}
				\caption{Scenario 1 - ego (blue) and lead (red) cars.}
				\label{fig:scen1_des}
\end{figure}
It is important to note that the last constraint in (\ref{eq:constraintsim}) ensures a safe distance between the ego vehicle and any vehicle that may suddenly appear in its path. This constraint plays a critical role in enhancing safety, particularly during dynamic scenarios such as overtaking, lane changes, or when a new obstacle enters the ego vehicle's lane unexpectedly. By enforcing this condition and by exploiting radar information, the controller is able to proactively reduce the risk of collisions through timely adjustments to the vehicle's trajectory and speed.
Furthermore note that, for the second scenario, the constraint on state $x_5$ in (\ref{eq:constraintsim}) is replaced by:
\begin{equation}
\left\{\begin{array}{lcl}
	-5 &\leq  x_5 \leq & 5 
\end{array}\right.
\label{eq:constraintsim2}
\end{equation}
The constraints on $x_5$ ensure safe lateral margins relative to the road geometry, distinguishing between the two-lane structure of Scenario 1 and the three-lane road of Scenario 2. Under these conditions, we evaluate control performance in two increasingly complex scenarios, assuming environmental awareness via V2X or on-board sensors. Figure \ref{fig:flow_chart} illustrates the control logic. Following initialization, the vehicle maintains cruise speed until a slower lead vehicle is detected, triggering a safe following mode. An overtaking maneuver is executed only upon the successful computation of a feasible trajectory; otherwise, the vehicle persists in safe following. This logic ensures that overtaking occurs only when valid planning guarantees safety, allowing the system to continuously adapt to dynamic traffic.

\begin{figure}[t!]
		\centering
		\includegraphics[angle=-90, width=1\linewidth]{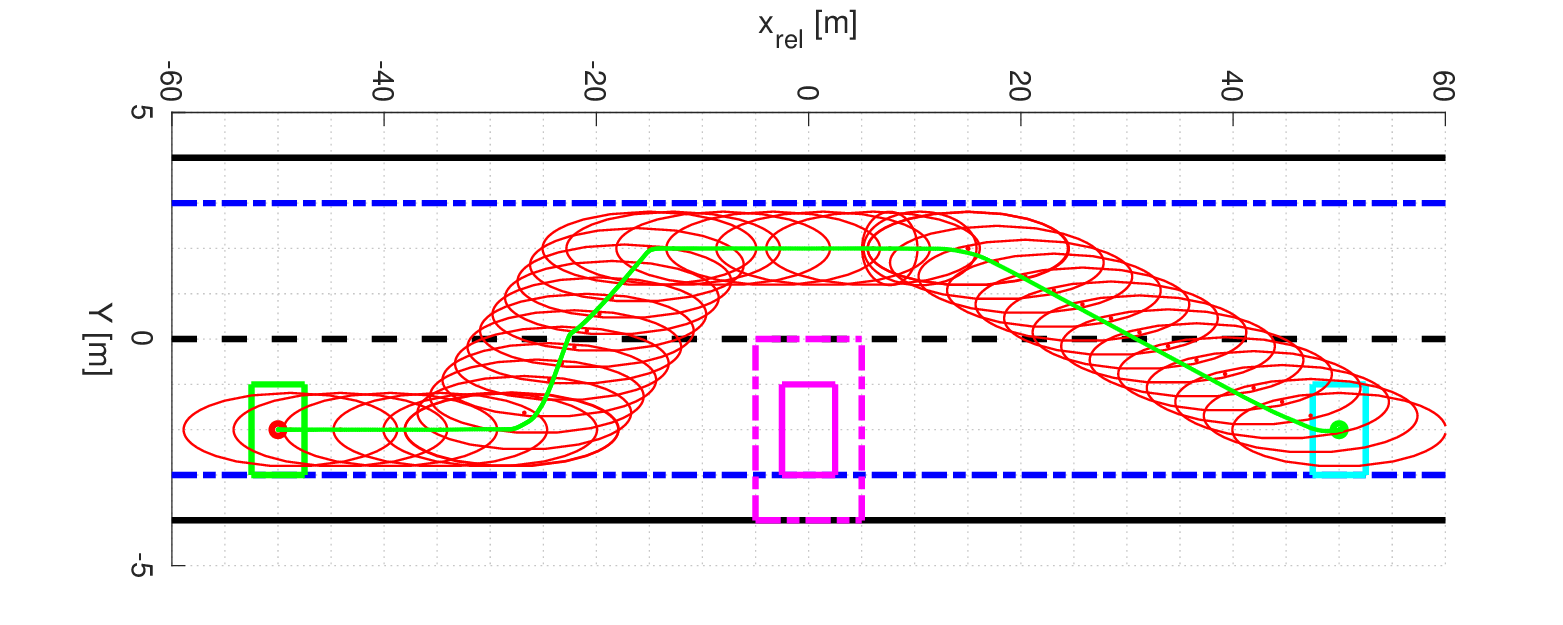}
				\caption{Scenario 1 - saturation region related to lateral position in the global reference frame ($x_5$) and relative distance ($x_6$), initial (red circle in the green rectangle) and final position (green circle in the cyan rectangle) of the ego car, lead car position (magenta rectangle), constraints on lateral position (dashed blue line), road lines (black lines and black dashed line) and ego vehicle trajectory (green line).}
				\label{fig:ocspscenario1}
\end{figure}
\subsection{Scenario 1: Standard Highway Overtaking}
\label{s6_1}
This scenario involves a two-lane highway where the ego vehicle overtakes a slower lead vehicle (Fig. \ref{fig:scen1_des}). The road consists of two $4\,$[m] lanes ($8\,$[m] total width), with the ego vehicle initially positioned at $Y = -2\,$[m] (center of the right lane). The lead vehicle acts as a dynamic obstacle traveling at a constant speed of $\dot{x}^{lead} = 20\,$[m/s] ($72\,$[km/h]). The maneuver initiates at a relative longitudinal distance of $x^{rel} = 50\,$[m], providing sufficient clearance for a safe transition to the left overtaking lane.

%This scenario represents a highway overtaking maneuver in a two-lane road with both lanes oriented in the same direction. The right lane is designated for normal driving, while the left lane is used for overtaking (Figure \ref{fig:scen1_des}). The lead vehicle travels at a constant cruising speed and represents a dynamic obstacle.
%
%The following configuration summarizes this first scenario:
%\begin{enumerate}
%	\item Road width: $8[m]$;
%	\item Lane width: $4[m]$;
%	\item Vehicle involved: ego vehicle and lead vehicle;
%	\item Lead vehicle speed: $\dot{x}^{lead}=20[m/s]=72[km/h]$;
%	\item  Initial relative distance between the vehicles: $x^{rel}=50[m]$ (i.e. an useful distance to start an overtaking maneuver);
%	\item  Ego vehicle lateral position: 2 meters from the centre-line ($Y=-2[m]$).
%\end{enumerate}

Based on the above scenario configuration, we define the initial state as
\begin{equation} x_0=\left[ \begin{array}{cccccc} 0 & 0 & 0 & 0 & -2 & -50 \end{array} \right]^T \label{eq:scen1x0} \end{equation}

whereas the final state is set as

\begin{equation} x_f=\left[ \begin{array}{cccccc} 0 & 0 & 0 & 0 & -2 & 50 \end{array} \right]^T \label{eq:scen1xf} \end{equation}
This configuration ensures that, after completing the overtaking maneuver, the ego vehicle is positioned 50 meters ahead of the lead vehicle.

%For the presented scenario, to create a domain of attraction containing the initial state $x_0$, we have computed 31 ellipsoidal families of 50 ellipsoids each. 
Figure \ref{fig:ocspscenario1} reports the saturation region related to lateral position in the global reference frame ($x_5$) and relative distance ($x_6$), the initial (red circle in the green rectangle) and final position (green circle in the cyan rectangle) of the ego car, the ego car position (green rectangle initial position, cyan rectangle final position), the lead car position (magenta rectangle), the constraints on lateral position (dashed blue line) and the road lines (black lines and black dashed line). Note also that the obstacle free region is so that the vehicles dimension are accounted as an obstacle. Consequently, the safe space, represented by the magenta dashed rectangle, is treated as an obstacle during the overtaking maneuver.

In this scenario, the ego vehicle travels straight until radar detects a sudden obstacle, triggering deceleration. Upon computing a feasible trajectory, it executes a left lane change to overtake and proceeds to the target.
\begin{figure}[h!]
		\centering
		\includegraphics[width=0.9\linewidth]{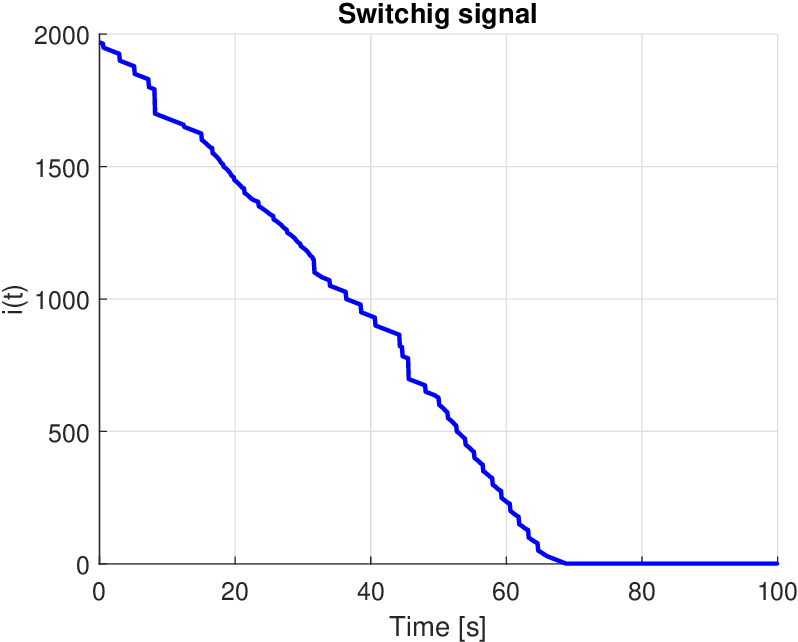}
		\caption{Scenario 1 - signal $i(t)$.}
		\label{fig:indice1}
\end{figure}
\begin{figure}[h!]
		\centering
		\includegraphics[width=0.9\linewidth]{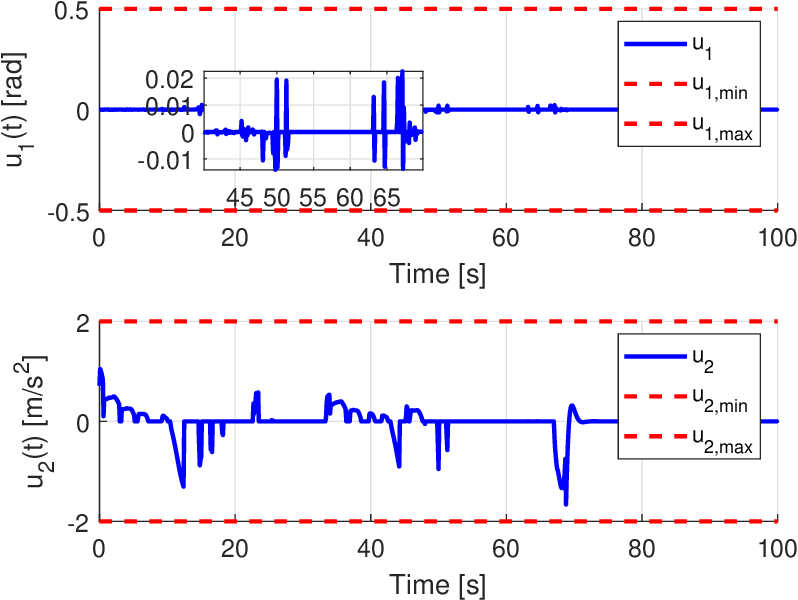}
		\caption{Scenario 1 - command inputs $u_1(t)$ and $u_2(t)$.}
		\label{fig:input1}
\end{figure}
\begin{figure}[h!]
		\centering
		\includegraphics[width=0.9\linewidth]{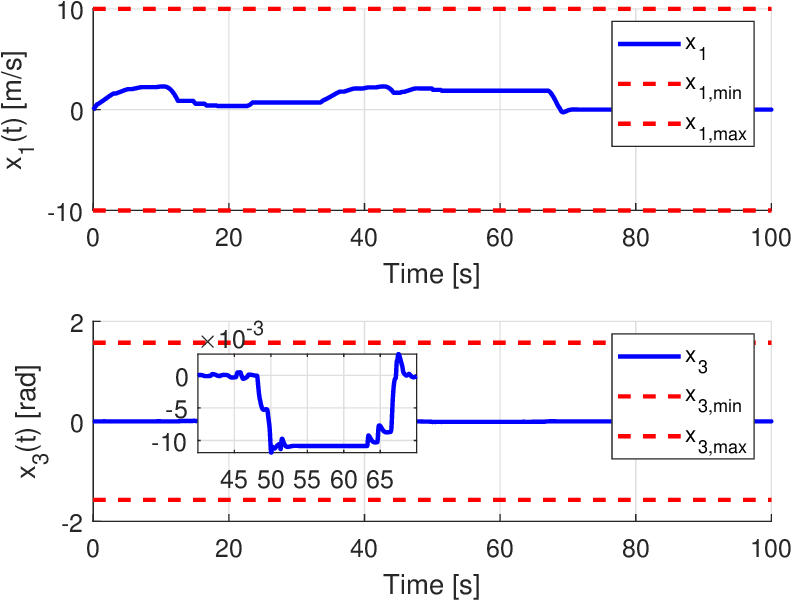}
		\caption{Scenario 1 - states ($x_1$ and $x_3$) evolution.}
				\label{fig:x1x31}
\end{figure}
\begin{figure}[h!]
		\centering
		\includegraphics[width=0.9\linewidth]{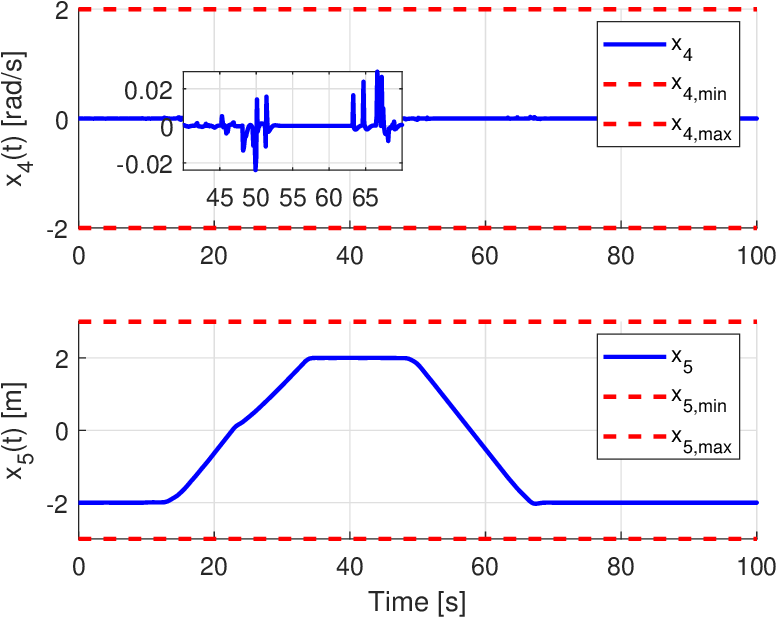}
			\caption{Scenario 1 - states ($x_4$ and $x_5$) evolution.}
			\label{fig:x4x51}
\end{figure}
All relevant results for this simulation scenario have been reported in Figures \ref{fig:indice1}-\ref{fig:x4x51}. In particular, Figure \ref{fig:indice1} illustrates the evolution of the signal $i(t)$, which represents the smallest ellipsoid that contains the current vehicle state $x(t)$. This signal provides insight into the contraction properties of the control algorithm, ensuring that the vehicle remains within a feasible and safe trajectory throughout the maneuver.
The strictly monotonic decrease of the switching signal empirically validates the recursive feasibility of the algorithm, ensuring that the vehicle remains within the safe invariant regions throughout the maneuver. Furthermore, it guarantees the finite-time completion of the overtaking sequence. The controller deterministically steers the vehicle from the outermost perturbed ellipsoid to the target invariant set, providing formal safety guarantees during execution.
From a control performance perspective, as illustrated in Fig. \ref{fig:input1}, the ST-FE-MPC generates smooth, chatter-free input profiles that strictly respect the predefined physical bounds (±0.5 rad for steering and ±2 m/s$^2$ for acceleration). This demonstrates that the proposed strategy is fully compliant with real-world actuator limitations and passenger comfort requirements, effectively eliminating aggressive control chattering.
On the other hand,  Figures \ref{fig:x1x31}-\ref{fig:x4x51} show the state ($x_1$, $x_3$, $x_4$ and $x_5$) evolution related to this realizations. 
It is important to note that, because the proposed framework allows to avoid the aggressive control saturation that typically affects standard optimization-based controllers, consequently, as shown in the state evolutions (Figures \ref{fig:x1x31} and \ref{fig:x4x51}), the vehicle performs a seamless lateral transition ($x_5$) with minimal yaw rate oscillations ($x_4$), ensuring passenger comfort and strict trajectory adherence.

Finally, the vehicle trajectory is reported in Figure \ref{fig:ocspscenario1} (green line). The presented results clearly demonstrate that the vehicle successfully performs a safe overtaking maneuver while ensuring compliance with the constraints given in (\ref{eq:constraintsim}).

%
%\begin{figure}[h!]
%		\centering
%		\includegraphics[width=0.9\linewidth]{x7_s1.eps}
%			\caption{Scenario 1 - states ($x_1$ and $x_7$) evolution. The green band highlights the simulation interval during which the ego car decreases its speed to avoid a collision with the second lead car.}
%			\label{fig:x1x71}
%\end{figure}
%
%
%%%
%\clearpage
%
\subsection{Scenario 2: Multi-Vehicle Dynamic Overtaking}
\label{s6_3}
\begin{figure}[h!]
		\centering
		\includegraphics[angle=-90, width=0.8\linewidth]{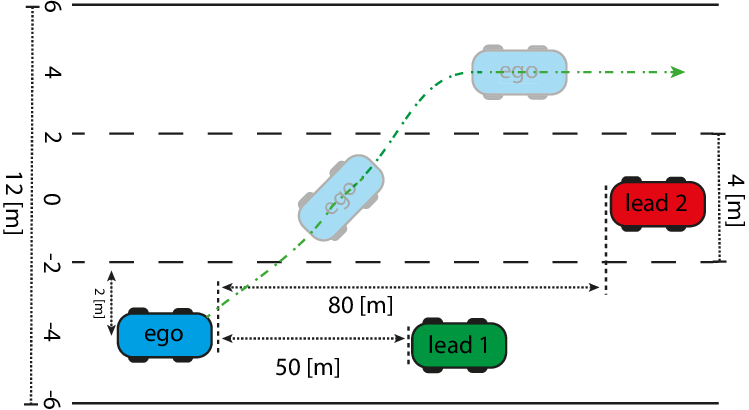}
				\caption{Scenario 2 - ego (blue) and lead (red and green) cars.}
				\label{fig:scen3_des}
\end{figure}
Scenario 2 increases complexity by introducing a three-lane road ($12\,$m total width) and a second dynamic obstacle to test real-time re-planning (Fig. \ref{fig:scen3_des}). Initially, the ego vehicle is centered in the right lane ($Y = -4\,$[m]), with lead car 1 ($Ob_1$) at $x^{rel} = 50\,$[m] and lead car 2 ($Ob_2$) in the middle lane at $x^{rel} = 80\,$m. While $Ob_1$ maintains a constant speed of $20\,$[m/s], $Ob_2$ exhibits dynamic variations, accelerating from $19\,$[m/s] to $20\,$[m/s] as it aligns with $Ob_1$. This setup forces the controller to adapt its trajectory mid-maneuver to account for the evolving obstacle configuration and varying relative velocities.
%
%Scenario 2 further increases the complexity by introducing dynamic speed variations and an obstacle configuration change during the maneuver. This tests the controller's ability to handle real-time scenario transitions while maintaining safety and efficiency.
%This scenario involves a straight, three-lane road with three vehicles: the ego car and two lead cars, all traveling in the same direction. Specifically, the ego car is positioned in the right lane, with lead car 1 located 50 meters ahead in the same lane. Meanwhile, lead car 2 is positioned in the middle lane, 80 meters ahead of the ego car (Figure \ref{fig:scen3_des}).

%The following configuration summarize the third scenario:
%\begin{enumerate}
%	\item The road has a total width of \( 12\,m \);
%	\item Each lane is \( 4\,m \) wide;
%	\item Two vehicles (the ego car and lead car 1) are positioned in the right lane, while lead car 2 %occupies the middle lane;
	%\item Lead car 1 travels at a constant cruising speed of \( \dot{x}^{lead}_1 = 20\,m/s \) (\( 72\,km%/h \));
%	\item Lead car 2 initially travels at a cruising speed of \( \dot{x}^{lead}_2 = 19\,m/s \) (\( 68.4%\,km/h \)), which increases to \( \dot{x}^{lead}_2 = 20\,m/s \) (\( 72\,km/h \)) when lead car 1 %aligns with it;
%	\item The relative distance between the ego car and lead car 1 is \( x^{rel} = 50\,m \);
%	\item The relative distance between the ego car and lead car 2 is \( x^{rel} = 80\,m \);
%	\item The ego vehicle is centered in the right lane at \( Y = -4\,m \).
%\end{enumerate}

Based on this scenario configuration, we define the initial state as:

\begin{equation} x_0=\left[ \begin{array}{cccccc} 0 & 0 & 0 & 0 & -4 & -40 \end{array} \right]^T \label{eq:scen3x0} \end{equation}

whereas the final state is set as:

\begin{equation} x_f=\left[ \begin{array}{cccccc} 0 & 0 & 0 & 0 & -4 & 50 \end{array} \right]^T \label{eq:scen3xf} \end{equation}
\begin{figure}[h!]
		\centering
		\includegraphics[angle=-90, width=1\linewidth]{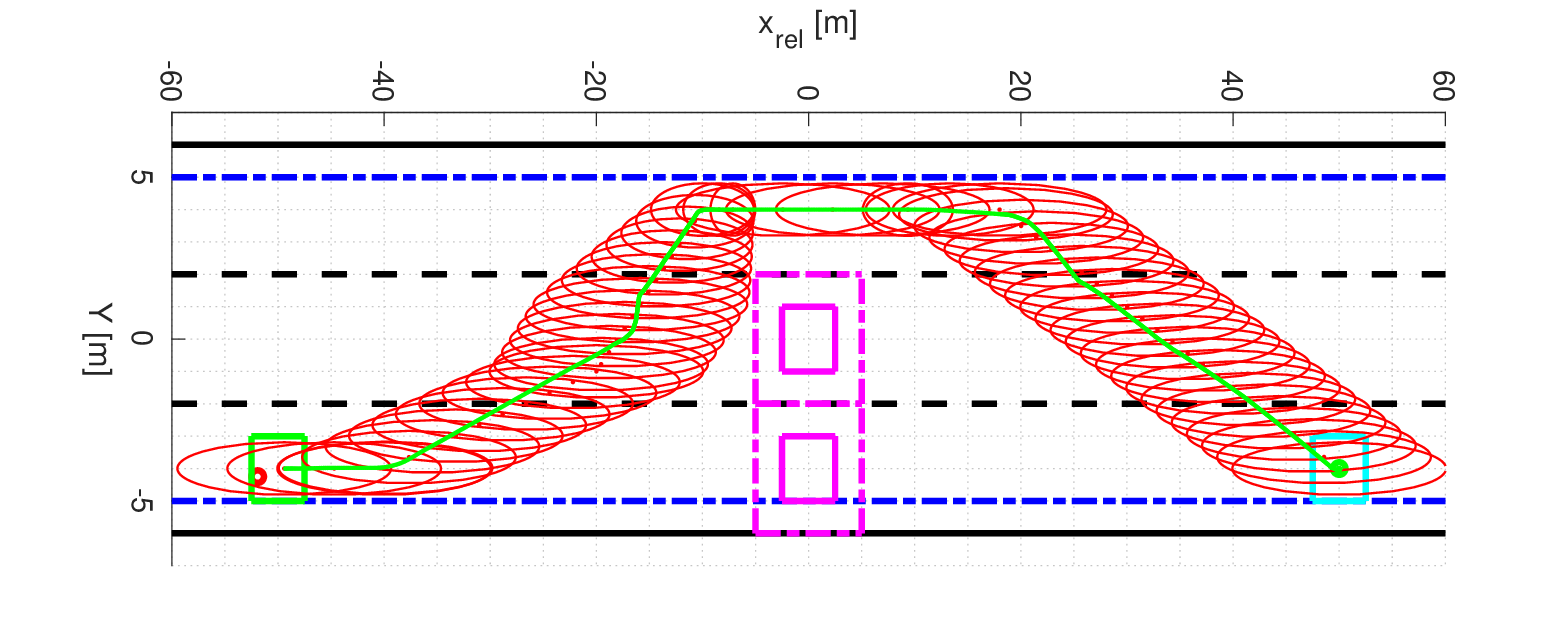}
				\caption{Scenario 2 - saturation region related to lateral position in the global reference frame ($x_5$) and relative distance ($x_6$), initial (red circle in the green rectangle) and final position (green circle in the cyan rectangle) of the ego car, lead cars position (magenta rectangle), constraints on lateral position (dashed blue line), road lines (black lines and black dashed line) and ego vehicle trajectory (yellow/green line).}
				\label{fig:ocspscenario3}
\end{figure}
This test account an obstacle scenario consisting of two object, i.e. lead cars 1 and 2. In particular this setting provide that the ego vehicle, to avoid collision with the lead car 1, starts an overtaking maneuver by shifting into the center lane by exploiting the feasible trajectory defined thanks to the sequence described by the red ellipsoid in Figure \ref{fig:ocspscenario3}. During this maneuver, while the ego vehicle is traveling in the center lane, the second vehicle appears in the same lane. The presence of this new car invalidates the originally planned trajectory. Consequently, the planned trajectory is updated so that the ego vehicle can shift into the leftmost lane, exploiting the feasible path defined by the sequence described by the red ellipsoids in Figure \ref{fig:ocspscenario3}. Moreover, as the radar sensor detects a decreasing distance to the vehicle ahead, the controller adaptively reduces the ego vehicle’s speed to preserve a safe following distance throughout the maneuver.

\begin{figure}[h!]
		\centering
		\includegraphics[width=0.9\linewidth]{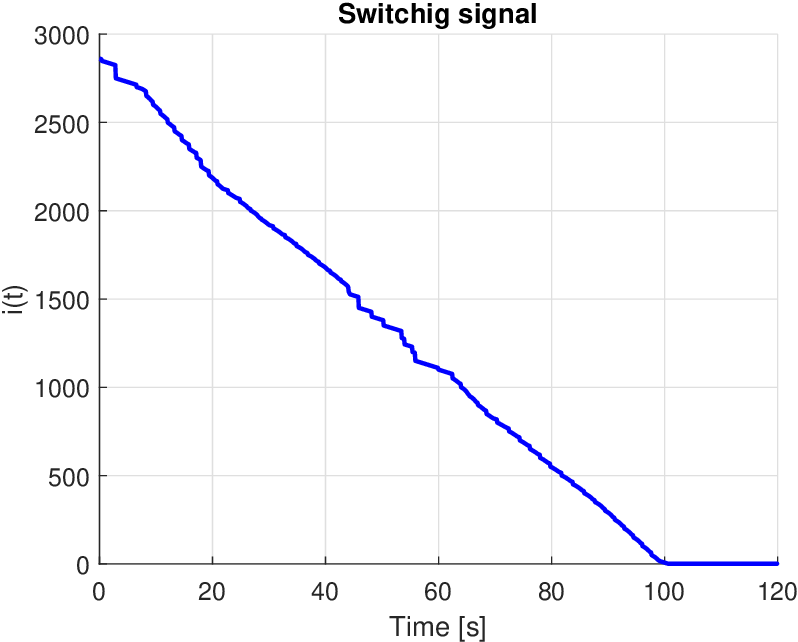}
		\caption{Scenario 2 - signal $i(t)$.}
		\label{fig:indice3}
\end{figure}

\begin{figure}[h!]
		\centering
		\includegraphics[width=0.9\linewidth]{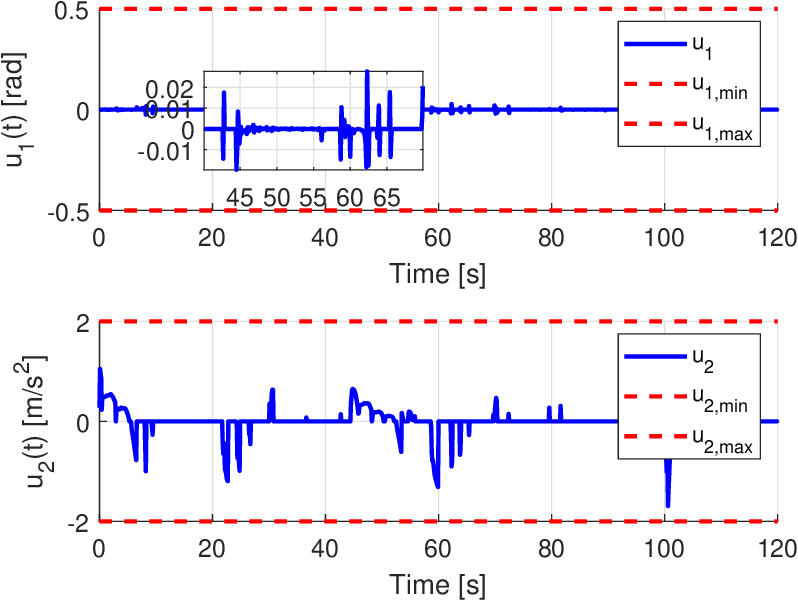}
		\caption{Scenario 2 - command inputs $u_1(t)$ and $u_2(t)$.}
		\label{fig:input3}
\end{figure}

\begin{figure}[h!]
		\centering
		\includegraphics[width=0.9\linewidth]{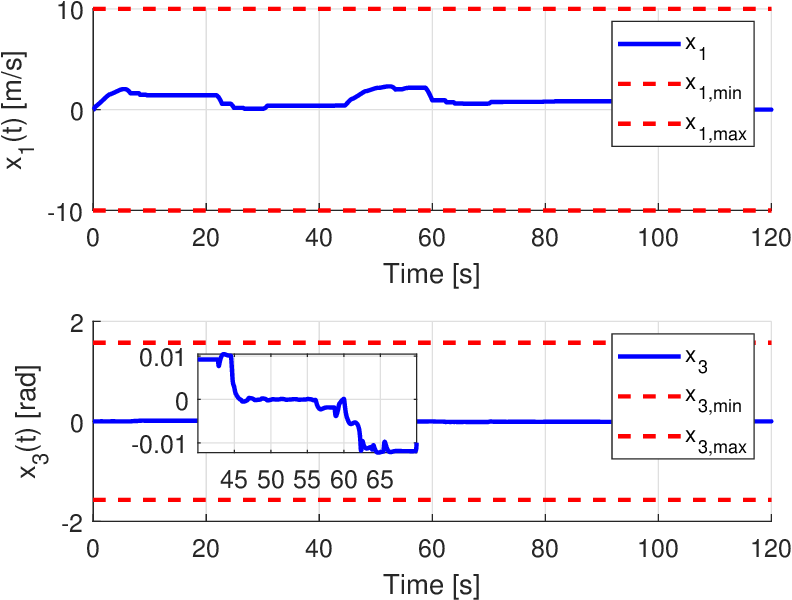}
		\caption{Scenario 2 - states ($x_1$ and $x_3$) evolution.}
				\label{fig:x1x33}
\end{figure}

\begin{figure}[h!]
		\centering
		\includegraphics[width=0.9\linewidth]{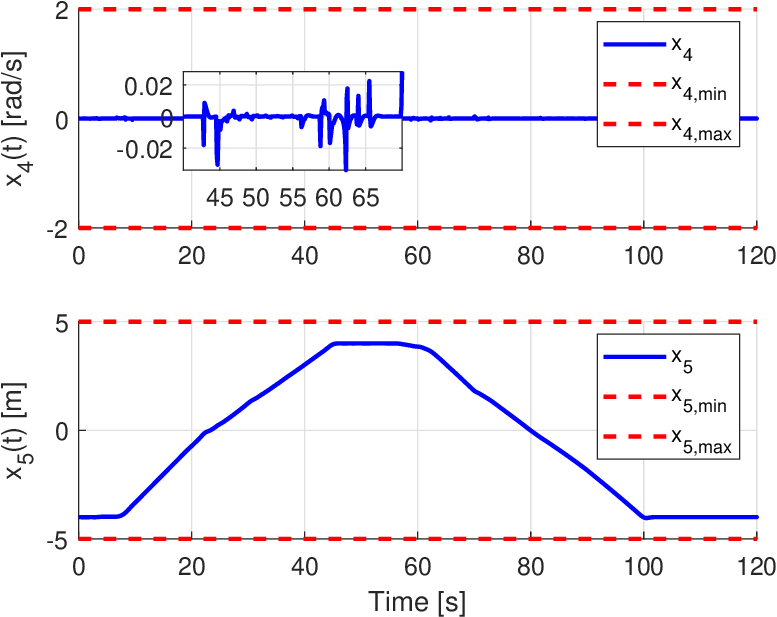}
			\caption{Scenario 2 - states ($x_4$ and $x_5$) evolution.}
			\label{fig:x4x53}
\end{figure}

\begin{figure}[h!]
		\centering
		\includegraphics[width=0.9\linewidth]{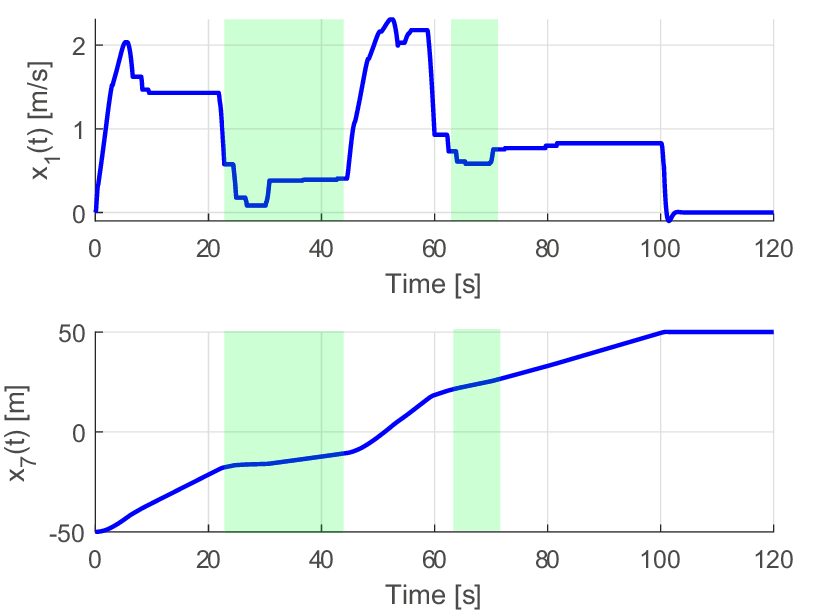}
			\caption{Scenario 2 - states ($x_1$ and $x_7$) evolution. The green band highlights the simulation interval during which the ego car decreases its speed to avoid a collision with the second lead car.}
			\label{fig:x1x73}
\end{figure}

All relevant results for this simulation scenario are presented in Figures \ref{fig:indice3}-\ref{fig:x1x73}. In particular, Figure \ref{fig:indice3} illustrates the signal \( i(t) \), which allows the analysis of the contraction level induced by the control algorithm throughout the system's evolution. Notably, this figure highlights the evolution of the  signal $i(t)$ across the sequences (red band pertain to the initial set sequence while the blue band refers to the update set sequence). 
Meanwhile, Figure \ref{fig:input3} presents the command inputs, while Figures \ref{fig:x1x33}-\ref{fig:x4x53} depict the state evolution of \( x_1 \), \( x_3 \), \( x_4 \), and \( x_5 \) for these realizations. Moreover, pay attention to Figure \ref{fig:x1x73}. This figure highlights the controller's capability to reduce the ego vehicle's speed in order to avoid a collision with two approaching vehicles in the same lane. In particular, this scenario requires the ego vehicle to execute a sequential overtaking maneuver: first avoiding lead car 1, and immediately evading a suddenly appearing lead car 2 while the initial action is still in progress.
As a result, the ego vehicle reduces its speed and shifts into the leftmost lane. 
Furthermore, this scenario serves as a quantitative evaluation of the system's robustness against the defined disturbance effects. When lead car 2 unexpectedly accelerates from 19 [m/s] to 20 [m/s], it explicitly induces a longitudinal disturbance $d_2(t) = \Delta v_x^{lead} \neq 0$ into the ego vehicle's relative dynamics. The performance shown in Figure \ref{fig:x1x33} demonstrates that the ST-FE-MPC algorithm seamlessly rejects this quantitative disturbance; the ego vehicle dynamically adjust its deceleration (Figure \ref{fig:input3}) to restore the target relative velocity  without violating the physical input constraints or breaching the minimum safe relative distance. 

Finally, the vehicle trajectory is shown in Figure \ref{fig:ocspscenario3}. 
The results clearly demonstrate that the vehicle successfully executes a safe overtaking maneuver while ensuring compliance with the constraints defined in (\ref{eq:constraintsim})\footnote{Simulation results obtained using the Automated Driving Toolbox (ADT) are available at the following link: \url{https://youtu.be/To5bNUW5gu8}.}.

%
%
%\subsection{Numerical Performance Evaluation}
%
%To assess the real-time feasibility of the proposed control algorithm, we conducted a series of computational experiments using MATLAB on a high-performance workstation equipped with an Intel® C621 chipset motherboard, dual Intel® Xeon® Gold processors running at 2.3 GHz (30 cores each), and 128 GB of DDR4 ECC RDIMM memory. The complete obstacle avoidance motion planning strategy %- including the offline computation of the terminal set and the online generation of one-step controllable sets- 
%was executed with varying complexity levels, primarily controlled by the maximum number of ellipsoids constructed per step. The algorithm was implemented using precompiled YALMIP optimizers interfaced with MOSEK, exploiting persistent caching mechanisms to minimize computational overhead.

%In a representative simulation scenario, the planned trajectory consisted of multiple waypoints aligned along a highway lane, with a inter-waypoint distance of approximately 10 meters. The computation time required to generate a feasible trajectory segment between two consecutive way-points averaged approximately 12 seconds. This average time includes the full construction of the ellipsoidal controllable sets and the corresponding local control synthesis. \hl{The observed performance confirms that, with appropriate 
%parameter tuning and  and , the proposed routine achieves update rates compatible with real-time constraints, supporting its applicability to embedded platforms or high-frequency control loops in autonomous vehicle systems.} 
\subsection{High-Fidelity Validation and Performance Evaluation}
\label{sec:carla_validation}
\begin{figure}[h]
\centering
\includegraphics[width=1\linewidth]{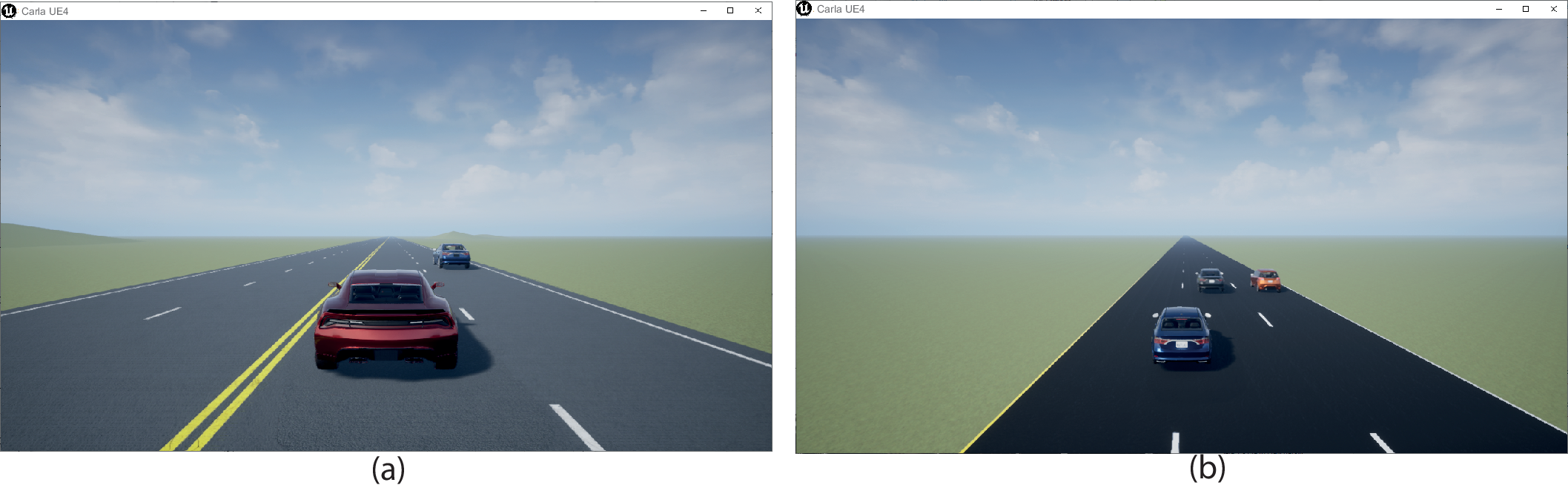}
\caption{CARLA Validation - Scenario 1 (a) and Scenario 2 (b).}
\label{fig:carla}
\end{figure}

\begin{figure}[h]
\centering
\includegraphics[width=0.9\linewidth]{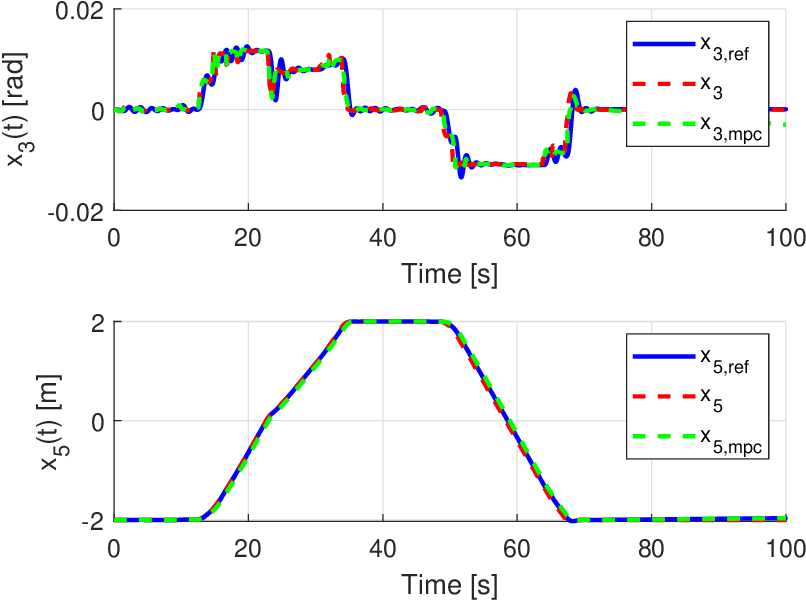}
\caption{Scenario 1 (CARLA) - Comparison: yaw angle ($x_3(t)$) and lateral position ($x_5(t)$). Blue: reference, red dashed: ST-FE-MPC, green dashed: NLMPC.}
\label{fig:xcomp_s1}
\end{figure}

\begin{figure}[h]
\centering
\includegraphics[width=0.9\linewidth]{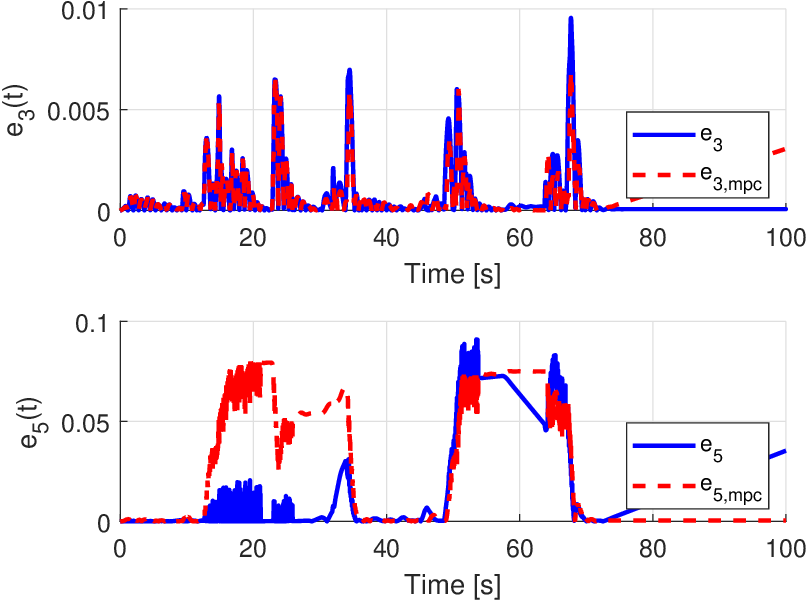}
\caption{Scenario 1 (CARLA) - Comparison: orientation ($e_3(t)$) and lateral ($e_5(t)$) errors. Blue: ST-FE-MPC, red dashed: NLMPC.}
\label{fig:ecomp_s1}
\end{figure}

\begin{figure}[h]
\centering
\includegraphics[width=0.9\linewidth]{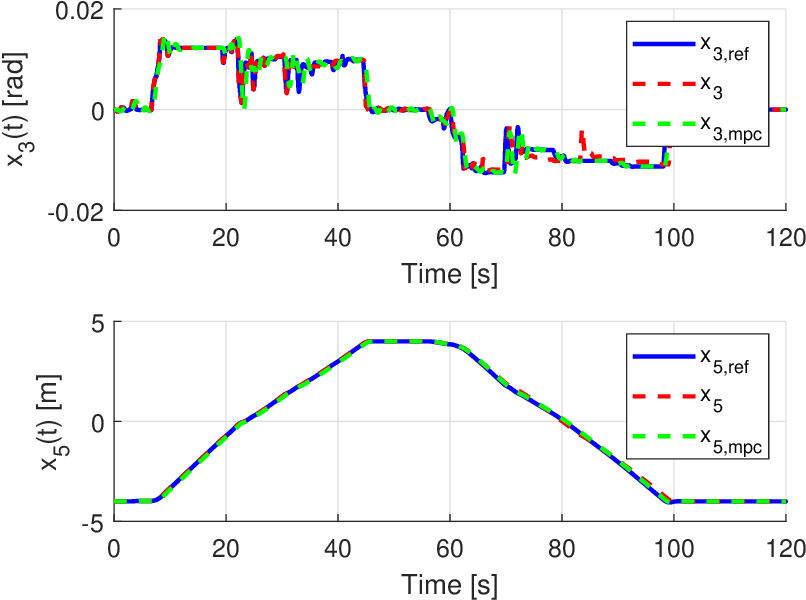}\caption{Scenario 2 (CARLA) - Comparison: yaw angle ($x_3(t)$) and lateral position ($x_5(t)$). Blue: reference, red dashed: ST-FE-MPC, green dashed: NLMPC.}
\label{fig:xcomp_s2}
\end{figure}

\begin{figure}[h]
\centering
\includegraphics[width=0.9\linewidth]{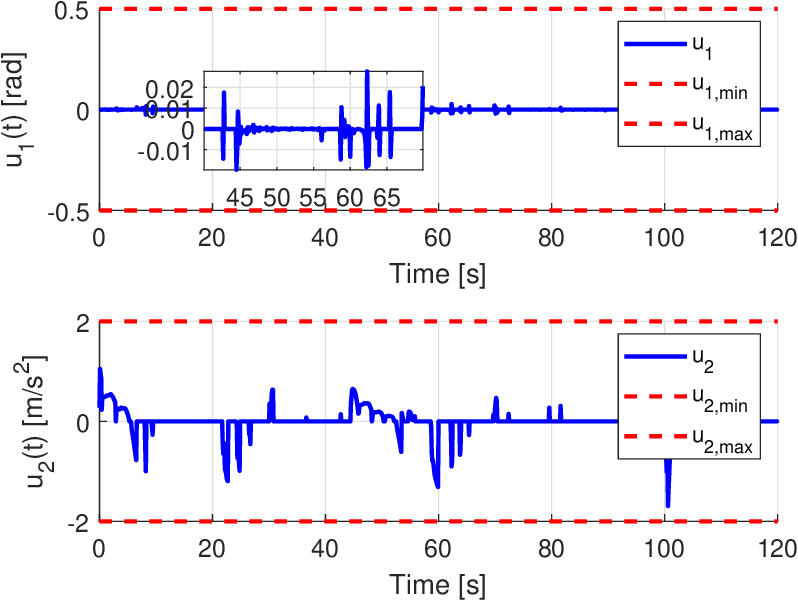}
\caption{Scenario 2 (CARLA) - Comparison: orientation ($e_3(t)$) and lateral ($e_5(t)$) errors. Blue: ST-FE-MPC, red dashed: NLMPC.}
\label{fig:ecomp_s2}
\end{figure}

To rigorously assess the practical viability and the engineering performance of the proposed ST-FE-MPC strategy beyond idealized mathematical environments, the control framework was validated using CARLA, a state-of-the-art open-source 3D simulation benchmark \cite{R5}. By interfacing the control logic with CARLA’s high-fidelity physical engine, the ego vehicle was exposed to realistic unmodeled dynamics, variable tire-road friction, and precise 3D spatial constraints (Fig. \ref{fig:carla}).
Within this environment, a comparative study was conducted against a baseline Non-Linear MPC (NLMPC) scheme across the dynamic maneuvers described in Scenarios 1 and 2. The controller's effectiveness is demonstrated across three critical domains:
\begin{itemize}
	\item Actuator Feasibility and Passenger Comfort: In both scenarios, the control inputs were strictly bounded to ensure stability. The steering angle $u_1(t)$ never exceeded ±0.5 rad, and the longitudinal acceleration $u_2(t)$ was maintained within [-2, 2] m/s², preventing aggressive saturation and ensuring smooth trajectories.
	\item Strict Safety Guarantees: The distance to leading vehicles ($x_6$) never dropped below the 12 m threshold, acting as a robust, mathematically certified Time-to-Collision (TTC) guarantee even under dynamic disturbances.
	\item Maneuver Efficiency: The operational efficiency is demonstrated by the rapid, monotonic decrease of the switching signal $i(t)$, guaranteeing maneuver completion within a mathematically bounded timeframe.
\end{itemize}
The comparative tracking performance is visually reported in Figs. \ref{fig:xcomp_s1}-\ref{fig:ecomp_s2}. 
The evolution of tracking errors (Figs. \ref{fig:ecomp_s1} and \ref{fig:ecomp_s2}) confirms convergence toward zero and robust disturbance rejection, with minimal transient peaks during dynamic disturbances in Scenario 2.
Moreover, to provide a quantitative evaluation, the Root Mean Square Error (RMSE) was computed:
\begin{equation}
\text{RMSE}_i = \sqrt{\frac{1}{T} \sum_{t=1}^{T} \| x_i(t) - \hat{x}_i(t) \|_2^2}, \quad i=1,\ldots,6
\label{rmse}
\end{equation}
As detailed in Table \ref{tab:table_rmse}, both controllers succeed in Scenario 1, but the performance gap widens in Scenario 2. Under unmodeled longitudinal disturbances, NLMPC exhibits significant degradation (lateral RMSE spikes to 0.4170), while ST-FE-MPC maintains tight accuracy (0.0969).
\begin{table}[h]
\centering
\caption{Root Mean Square Error (RMSE) for System States in CARLA}
\label{tab:table_rmse}
\begin{tabular}{lcccc}
\hline
\textbf{State} & \multicolumn{2}{c}{\textbf{Scenario 1}} & \multicolumn{2}{c}{\textbf{Scenario 2}} \\
 & \textbf{ST-FE-MPC} & \textbf{NLMPC} & \textbf{ST-FE-MPC} & \textbf{NLMPC} \\ 
\hline
1 & 0.0160 & 0.0450 & 0.0084 & 0.0092 \\
2 & 0.0250 & 0.0640 & 0.0074 & 0.0130 \\
3 & 0.0016 & 0.0015 & 0.0016 & 0.0016 \\
4 & 0.0290 & 0.0340 & 0.0140 & 0.0170 \\
5 & 0.0299 & 0.0398 & 0.0969 & 0.4170 \\
6 & 0.0430 & 0.0690 & 0.0054 & 0.0064 \\
\hline
\end{tabular}
\end{table}
Beyond precision, the core bottleneck is the online computational burden. Table \ref{tab:comp_comparison} reports average execution times for $T_s = 100$ ms.
\begin{table}[h]
\centering
\caption{Computational Performance Comparison ($T_s = 100$ ms)}
\label{tab:comp_comparison}
\begin{tabular}{llcc}
\hline
\textbf{Scenario} & \textbf{Controller} & \textbf{Average Time/Step} & \textbf{Real-Time Feasible} \\ \hline
\multirow{2}{*}{Scenario 1} & NLMPC & 108.33 ms & No ($> T_s$) \\
 & \textbf{ST-FE-MPC} & \textbf{7.12 ms} & \textbf{Yes (7.1\% of $T_s$)} \\ \hline
\multirow{2}{*}{Scenario 2} & NLMPC & 116.54 ms & No ($> T_s$) \\
 & \textbf{ST-FE-MPC} & \textbf{8.86 ms} & \textbf{Yes (8.9\% of $T_s$)} \\ \hline
\end{tabular}
\end{table}
NLMPC violates real-time constraints in both scenarios. In contrast, ST-FE-MPC consistently requires less than 9\% of the sampling window, proving its superior efficiency and deterministic real-time feasibility. 
The ST-FE-MPC structurally absorbs unmodeled variations, successfully confining the vehicle's state within safe invariant tubes and neutralizing the drift typical of standard linear controllers in complex traffic.

\section{Conclusions}
\label{s7}
This work introduced a set-theoretic Receding Horizon Control (RHC) strategy for autonomous vehicle motion planning and overtaking in highway scenarios. By leveraging inner ellipsoidal approximations of reachability sets for uncertain polytopic systems, the framework ensures safety, disturbance robustness, and strict constraint satisfaction. Validation in MATLAB/Simulink and CARLA confirmed the method's computational efficiency and adherence to engineering performance metrics.
Specifically, the approach guarantees stability by bounding lateral and longitudinal accelerations and maintains deterministic safe-following distances, ensuring maneuvers are both mathematically optimal and physically comfortable for real-world automotive applications.
While managing severe actuator failures falls outside the immediate scope of this OAMP framework, the proposed polytopic uncertainty structure inherently absorbs bounded parameter variations, providing robustness against mild mechanical wear. Future work will focus on extending the methodology to complex urban intersections, integrating machine learning for adversarial driving, and incorporating explicit Fault-Tolerant Control (FTC) mechanisms to address severe hardware degradation.

\section{Acknowledgment}
This work was partially supported by the European Union under the Italian National Recovery and Resilience Plan (NRRP) of NextGenerationEU, partnership on "Technologies for Climate Change Adaptation and Quality of Life Improvement" (ECS00000009 - program "TECH4YOU").

\end{document}